\documentclass[%
 reprint,
 amsmath,amssymb,
 aps,
]{revtex4-1}

\usepackage{graphicx}
\usepackage{dcolumn}
\usepackage{bm}
\usepackage{multirow}
\usepackage{color}
\usepackage{braket}

\newcommand{\Yb}{\ensuremath{^{171}\mathrm{Yb}^+~}}

\begin{document}


\title{Double-EIT Ground-State Cooling of Stationary Two-Dimensional Ion Lattices}
\author{Mu Qiao$^{1}$}
\email{mu.q.phys@gmail.com}
\author{Ye Wang$^{1,2}$}
\author{Zhengyang Cai$^{1}$}
\author{Botao Du$^{1,4}$}
\author{Pengfei Wang$^{1}$}
\author{Chunyang Luan$^1$}
\author{Wentao Chen$^1$}
\author{Heung-Ryoul Noh${^3}$}%
\author{Kihwan Kim$^{1}$}%
\email{kimkihwan@mail.tsinghua.edu.cn }
\affiliation{%
$^{1}$ Center for Quantum Information, Institute for Interdisciplinary Information Sciences, Tsinghua University, Beijing 100084, P. R. China\\
$^{2}$ Department of Electrical and Computer Engineering, Duke University, Durham, NC 27708, USA\\
$^{3}$ Department of Physics, Chonnam National University, Gwangju, 61186, Korea \\
$^{4}$ Department of Physics and Astronomy, Purdue University, West Lafayette, IN 47907, USA}


\date{\today}

\begin{abstract}
We theoretically and experimentally investigate double electromagnetically induced transparency (double-EIT) cooling of two-dimensional ion crystals confined in a Paul trap. The double-EIT ground-state cooling is observed for \Yb ions with clock state, for which EIT cooling has not been realized like many other ions with a simple $\Lambda$-scheme. A cooling rate of $\dot{\bar n}=34~(\pm1.8)~\rm{ms}^{-1}$ and a cooling limit of $\bar n=0.06~(\pm 0.059)$ are observed for a single ion. The measured cooling rate and limit are consistent with theoretical predictions. We apply double-EIT cooling to the transverse modes of two-dimensional (2D) crystals with up to 12 ions. In our 2D crystals, the micromotion and the transverse mode directions are perpendicular, which makes them decoupled. Therefore, the cooling on transverse modes is not disturbed by micromotion, which is confirmed in our experiment. For the center of mass mode of a 12 ions crystal, we observe a cooling rate and a cooling limit that are consistent with those of a single ion, including heating rates proportional to the number of ions. This method can be extended to other hyperfine qubits, and near ground-state cooling of stationary 2D crystals with large numbers of ions may advance the field of quantum information sciences.
\end{abstract}

\maketitle

Cooling down mechanical oscillators into their ground states facilitates experimental investigations and applications with atoms and ions for quantum information sciences \cite{Wineland2013Novel}. Quantized vibrations of mechanical oscillators can be used as resources for continuous-variable quantum computation \cite{Lloyd1999Quantum,Lau2016Universal,Ding2017Quantum,Ding2017Cross,fluhmann2019encoding} or Boson sampling \cite{Aaronson2011,Lau2012Proposal,shen2014scalable,toyoda2015hon,um2016phonon,shen2018quantum}, which begins with ground state preparation. In order to demonstrate quantum advantages with these applications, ground-state cooling dozens of vibrational modes is required \cite{Aaronson2011}.

The performance of quantum operations with atoms and ions can be improved by ground-state cooling of vibrational degrees of freedom. Thermally induced phase noise and amplitude fluctuations of qubit-qubit interaction can also be suppressed by ground-state cooling, which is essential for realizing high-fidelity quantum gates \cite{ballance2016high,harty2016high} and reliable quantum simulations \cite{blatt2012quantum,monroe2019programmable}. Moreover, quantum simulations with both vibrational and fermionic degrees of freedom \cite{mezzacapo2012digital,lv2018quantum} naturally demand ground-state cooling of vibrational modes. As the sizes of quantum systems scale up, efficient ground-state cooling for large numbers of modes becomes even more necessary for high-fidelity quantum manipulations.

By removing the entropy from oscillators to photons, laser cooling provides a practical way to prepare ground state of atomic and even macroscopic oscillators. Laser cooling was first experimentally demonstrated by velocity-dependent radiative force  \cite{wineland1978radiation,neuhauser1978optical}, which is known as Doppler cooling. The final temperature can be reached with Doppler cooling is limited by the natural line-width of the cooled atoms. Sisyphus cooling provides lower temperature than the Doppler limit \cite{dalibard1989laser,ejtemaee20173d},  which has been widely used with neutral atoms. Recently, it was found that Sisyphus cooling can also be applied to trapped ions \cite{ejtemaee20173d}. Although ground-state cooling can be realized by resolved-sideband cooling \cite{monroe1995resolved, roos1999quantum}, the narrow excitation range of resolved-sideband transitions makes it difficult to perform simultaneous ground-state cooling for multiple motional modes of large crystals. Moreover, some of sideband transitions driven by high power lasers induce a charging problem \cite{harlander2010trapped}, which is even worse for UV laser beams.

\begin{figure*}[!htb]
\center{\includegraphics[width=0.9\textwidth]{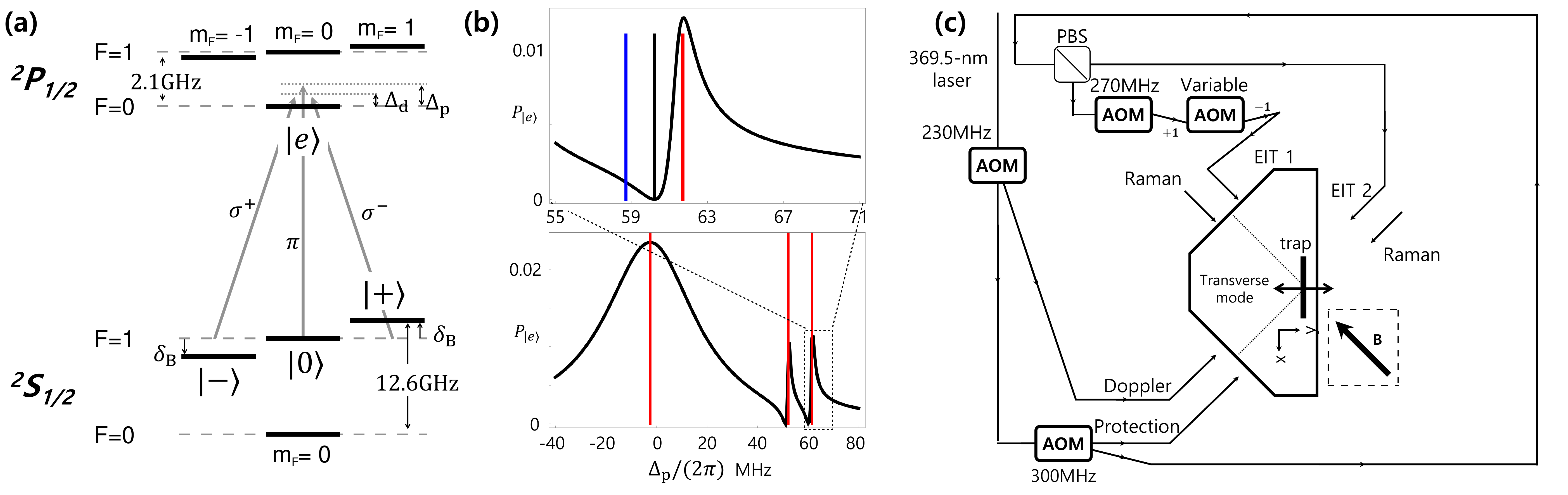}}
\caption{\label{fig:setup}(a)Relevant energy levels of \Yb for EIT cooling. (b)Fano-like profile of double-EIT. The spectrum is calculated by steady state solution of the master equation or by scattering amplitude \cite{supple}. In the simulation we set $\Delta_{\rm{d}}/(2\pi)=55.6$MHz, $\delta_{\rm{B}}/(2\pi)=4.6$MHz, $\Omega_{\sigma,\pm}/(2\pi)=17$MHz, $\Omega_\pi/(2\pi)=4$MHz, $\nu/(2\pi)=1.5$MHz, where $\nu$ is the frequency of the mode we intend to cool down. The bottom one shows the spectrum in a large range while the red lines represent theoretical predictions of the positions of dressed states. The top one shows the spectrum around the peak we use for cooling, and the blue (red) line represents the position of motional sideband, while the dark line represents carrier transition. The red-sideband has lower energy, which corresponds to a higher detuning. (c) Optical configuration. The EIT beam is first separated from the Doppler cooling beam with 14 GHz sideband, then is split into the driving beam and the probe beam by a PBS. The relative detuning, $\Delta_{\rm d}-\Delta_{\rm p}$, are controlled by two AOMs acted on the driving beam. The first-order diffraction of the 270MHz AOM and the negative first-order of the variable AOM is used. The net propagating vectors $\Delta k$ of both the EIT and the Raman beams are along the direction of transverse mode. A quarter-wave plate (QWP) is used to adjust the polarization of the driving beam.
}
\end{figure*}

Electromagnetically-Induced-Transparency (EIT) cooling \cite{morigi2000ground,roos2000experimental,lin2013sympathetic} provides an alternative possibility: It can apace cool down a wide range of vibrational modes simultaneously, which has been demonstrated in the linear trap and the Penning trap with tens to hundreds of ions \cite{lechner2016electromagnetically,jordan2019near}. Typical EIT cooling uses quantum interference in a three-level $\Lambda$-scheme, and has only been implemented for ions without clock states. Here, we demonstrate a novel cooling method for \Yb ions with clock state, based on double-EIT \cite{paspalakis2002transparency, beck2017propagation, alotaibi2014double, ham2000coherence,lee2014experimental,wang2017strong} in a four-level system. Double-EIT cooling has been theoretically studied \cite{evers2004double,yi2013ground,huang2016double,semerikov2018eit}, and a variant of it has been implemented with $^{40}\rm{Ca}^+$ ion \cite{scharnhorst2018experimental}. We experimentally perform double-EIT cooling of \Yb ions to prepare motional ground states of a two-dimensional (2D) ion crystal. We cool down the transverse modes perpendicular to the crystal plane in which the micromotion oscillates \cite{wang2019realization}, therefore the cooling efficiency is negligibly affected by micromotion. The efficiency of double-EIT cooling is systematically studied as a function of various control parameters including the intensity and the detuning of the probe and the driving laser beams to obtain optimal conditions. For multiple motional modes, crystals are cooled down near to their ground states in hundreds of microseconds with cooling rates similar to that of a single ion. 

Double-EIT cooling for \Yb ions involves four energy levels, which is different from EIT cooling in a three-level $\Lambda$ scheme. As shown in Fig.\ref{fig:setup} (a), the excited-state $\ket{e}\equiv |F=0,m=0\rangle$ in $P_{\frac{1}{2}}$ manifold is coupled to three states of $\ket{-}\equiv\ket{F=1,m=-1}$, $\ket{0}\equiv\ket{F=1,m=0}$, and $\ket{+}\equiv\ket{F=1,m=+1}$ in $S_{1/2}$ manifold. The four-level system can be regarded  as two of $\Lambda$-schemes, which produces two Fano-like profiles in the absorption spectrum. For instance, one of the $\Lambda$ schemes consists of the $\ket{-}$, $\ket{0}$ (or $\ket{+}$, $\ket{0}$) states and the excited state $\ket{e}$, which are coupled by the driving beam with $\sigma^{+}$ (or $\sigma^{-}$) polarization and the probe beam with ${\pi}$ polarization, respectively. As shown in the Fig.\ref{fig:setup}(b), the absorption spectrum of the probe beam for an ion at rest has two null points corresponding to two-dark states when the detuning of the transition $|0\rangle\leftrightarrow |e\rangle$ matches the detuning of the transitions $|\pm\rangle\leftrightarrow |e\rangle$ \cite{supple}. And the two narrow peaks correspond to dressed states formed by $|\pm\rangle$ and $|e\rangle$ \cite{supple}. We determine the distances between null points and the corresponding narrow peaks by the ac Stark shift of dressed states. 

The principle of double-EIT cooling is similar to that of single-EIT cooling, which uses the asymmetric absorption profile to enhance red-sideband transitions and suppress carrier/blue-sideband transitions, as shown in Fig. \ref{fig:setup}(b). The broad width of the peak enables wide-range cooling. The motional modes of large crystals can be efficiently cooled down to near ground state based on the unbalanced scattering amplitude between red- and blue-sideband transitions. When the detuning $\Delta_{\rm {p}}$ of the probe beam is set equal to $\Delta_{\sigma^{+}}\equiv\Delta_{\rm d}+\delta_{\rm B}$, the internal state of ion is pumped to a dark state, and the ion will not absorb any photon unless the ion motion induces a differential Doppler shift ${\vec{v}\cdot(\hat k_\pi-\hat k_{\sigma^{+}})}/{c}=\delta_{+}$ between the $\pi$ and $\sigma^{+}$ transitions. As stated above, double-EIT can only cool down motional modes non-perpendicular to the difference in wave vector ($\hat k_\pi-\hat k_{\sigma^{+}}$). Therefore, the net $k$-vector should be aligned to the direction of the motional modes of interest. In our experiment, we choose the right peak for cooling; however, both peaks in the absorption spectrum can be used with similar cooling rates and limits. In principle, it is possible to make only one peak dominant, similar to the simple $\Lambda$-system, by unbalancing the Rabi frequencies of the $\sigma^{+}$ and $\sigma^{-}$ components of the driving beam. However, we do not observe an enhancement of cooling efficiency by using an unbalanced driving beam.

We experimentally demonstrate double-EIT cooling with \Yb ions, which have a clock-state qubit with a coherence time of over 10 minutes \cite{wang2017single}. The energy splitting $\omega_{0}$ of the qubit states $|F=1,m=0\rangle$ and $|F=0,m=0\rangle$ in $S_{\frac{1}{2}}$ manifold is 12.642812 GHz. The \Yb ions are trapped in a pancake-like potential produced by a radio-frequency Paul trap as described in Ref. \cite{wang2019realization}, where trapped ions can form a 2D crystal. A B-field of 3.32 Gauss is horizontally applied to break the dark state resonance in Doppler cooling, as shown in Fig.\ref{fig:setup} (c).

The EIT beams consist of two lasers, which are close to the $S_{\frac{1}{2}}|F=1,m=0\rangle$ to $P_{\frac{1}{2}}|F=0,m=0\rangle$ transition. The EIT beams are aligned to make the difference in wave vectors parallel to the transverse direction of motional modes. One of the beams serves as driving the $\sigma_{\pm}$ transitions between $\ket{\pm}\leftrightarrow\ket{e}$. The other beam works as probe beam, which couples the energy levels $\ket{0}\leftrightarrow\ket{e}$. The detuning $\Delta_{\rm p}$ of the probe beam is fixed at $(2\pi)$55.6 MHz, and the detuning $\Delta_{\rm d}$ of the driving beam is adjusted by altering frequency difference between two AOMs, as shown in Fig. \ref{fig:setup}(c). We measure the Rabi frequency and the polarization of the EIT beams by observing the differential ac Stark shift of the clock state qubit and the Zeeman state qubits \cite{haffner2003precision,ejtemaee20173d,supple}. The Rabi frequencies, $\{\Omega_{\sigma_-},\Omega_{\pi},\Omega_{\sigma_+}\}/(2\pi)$, of the driving beam of 24$\mu$W and the probe beam of 5.5$\mu$W are \{16.74,1.72, 18.03\}MHz  and \{1.49, 6.67, 3.17\}MHZ. 

\begin{figure}[!htb]
\center{\includegraphics[width=0.5\textwidth]{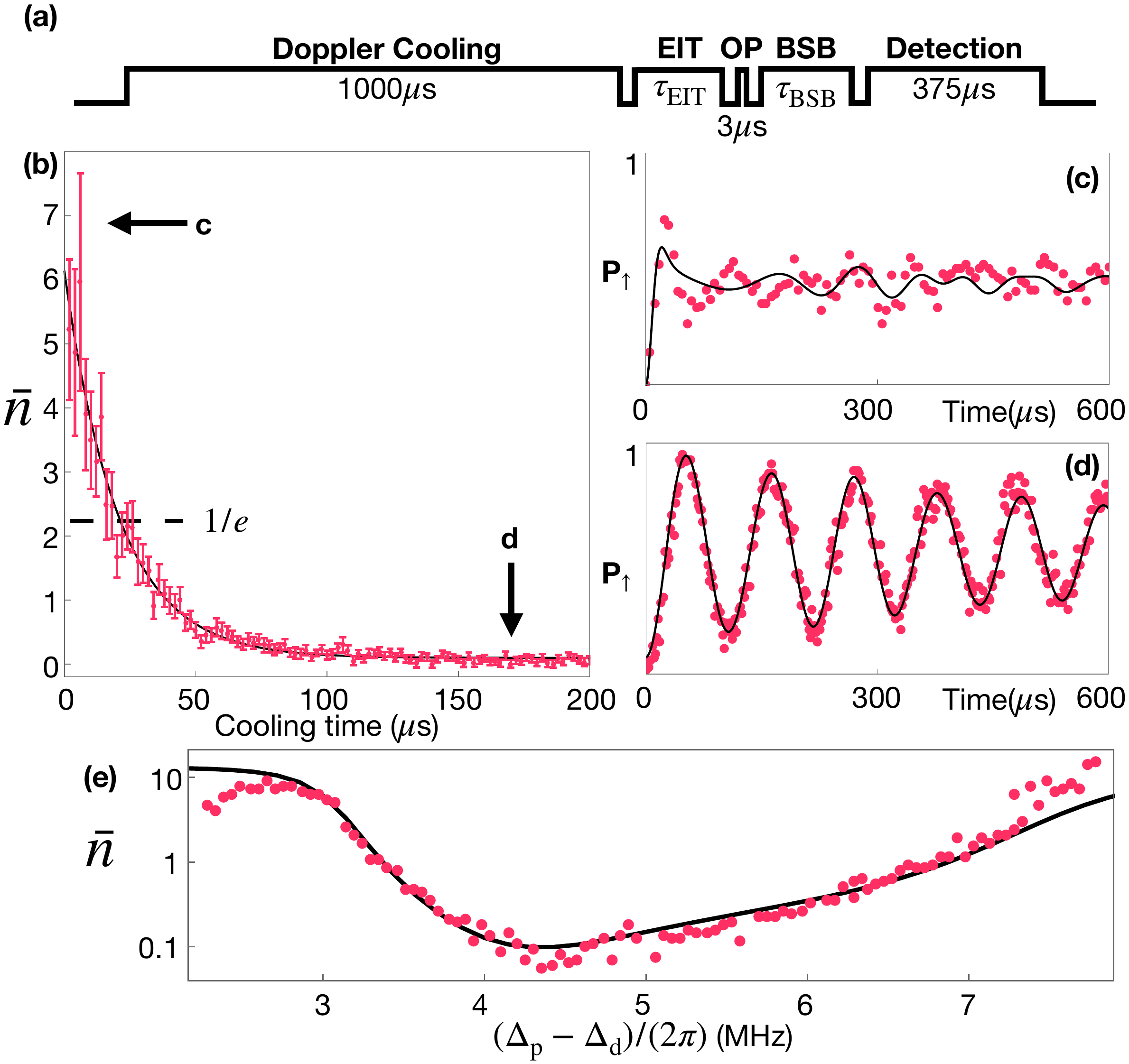}}
\caption{\label{fig:speed}(a) Experimental sequence for exploring EIT cooling of a single trapped ion. (b) Cooling dynamics for the transverse mode along the y-axis. Red points are experimental data obtained by fitting blue-sideband transitions shown in (c) and (d). Error bars denote fitting errors. The black line is exponential fit. The horizontal dashed line indicates 1/e of initial phonon number. (c,d) The blue-sideband transition after (c) Doppler cooling and (d) EIT cooling of 200$\mu$s. (e) Average phonon number $\bar{n}$ at the end of double-EIT cooling versus the relative detuning between the probe beam and the driving beam. The black line is numerical simulation result obtained by solving the master equation \cite{supple}.}
\end{figure}

Fig.\ref{fig:speed} (a) shows the experimental sequence to study double-EIT cooling with a single ion. For a single \Yb ion, secular trap frequencies are $\omega_{\rm y}/2\pi=2.38 \text{ MHz}$ in the transverse direction and $\{\omega_{\rm x},\omega_{\rm z}\}/2\pi=\{0.42,0.47\}\text{ MHz}$ in the crystal plane. We first apply Doppler cooling, which leads to the Doppler-limit temperatures around phonon number $\bar{n}\approx 7$. After Doppler cooling, 95\% population of the internal state of ions falls into the $S_{\frac{1}{2}}|F=1\rangle$ manifold. Afterwards, we apply the EIT beams for a duration $\tau_{\rm{EIT}}$. In order to measure the final phonon number $\bar n$, 3$\mu$s optical pumping is carried out to prepare the ground state $S_{\frac{1}{2}}|F=0,m=0\rangle$. By driving blue-sideband transition and fitting time evolution \cite{leibfried2003quantum}, the average phonon number $\bar n$ is extracted.

We experimentally study double-EIT cooling dynamics, with relative detuning $\Delta_{\rm p}-\Delta_{\rm d}=4.55\rm{MHz}$, by measuring the mean occupation number $\bar n$ at various cooling duration $\tau_{\rm EIT}$, as indicated in Fig.\ref{fig:speed} (b). The mean vibrational number $\bar n$ is measured by fitting the blue-sideband transitions, which are shown in Fig.\ref{fig:speed} (c) before and (d) after EIT cooling. Without EIT cooling, oscillations on the blue-sideband transition decay fast due to various excitations on different vibrational number states with different Rabi frequencies. As shown in Fig.\ref{fig:speed} (d), the minimum value of $\bar n_{\min}=0.06(\pm 0.059)$ of EIT cooling demonstrates a near ground-state cooling similar to sideband cooling. The $1/e$ cooling time $\tau_{\rm cool}=1/\gamma_{\rm cool}$, where $\gamma_{\rm cool}$ is cooling rate, is $30(\pm 1.6)\mu$s. A duration of 200 $\mu$s is sufficient to reach ground state.

By changing the frequency difference between the EIT beams, we determine the cooling range and the optimal detuning for double-EIT cooling. The efficiency of EIT cooling is determined by the ratio of absorption strengths between red-sideband and blue-sideband transitions, as shown in Fig. \ref{fig:setup}(b), which is controlled by the detuning of the driving beam $\Delta_{\rm d}$ in our experiment. The optimal detuning $(\Delta_{\rm p}-\Delta_{\rm d})/(2\pi)$ for double-EIT cooling locates at 4.55 MHz. This value is in accordance with the predicted value of 4.57 MHz, which can be calculated by $\delta_{\rm B}+\delta_{\rm DR}-\nu$, where $\delta_{\rm DR}$= (2$\pi$) 2.31 MHz is dressed-state ac Stark shift \cite{supple}.  Numerical simulations are performed to assess the experimental results, including a heating rate of $0.67\rm{ms}^{-1}$ along the transverse direction. The solid line in Fig.\ref{fig:speed} (e) indicates the simulated average phonon numbers, which match the experimental results fairly well.

\begin{figure}[!htb]
\center{\includegraphics[width=0.5\textwidth]{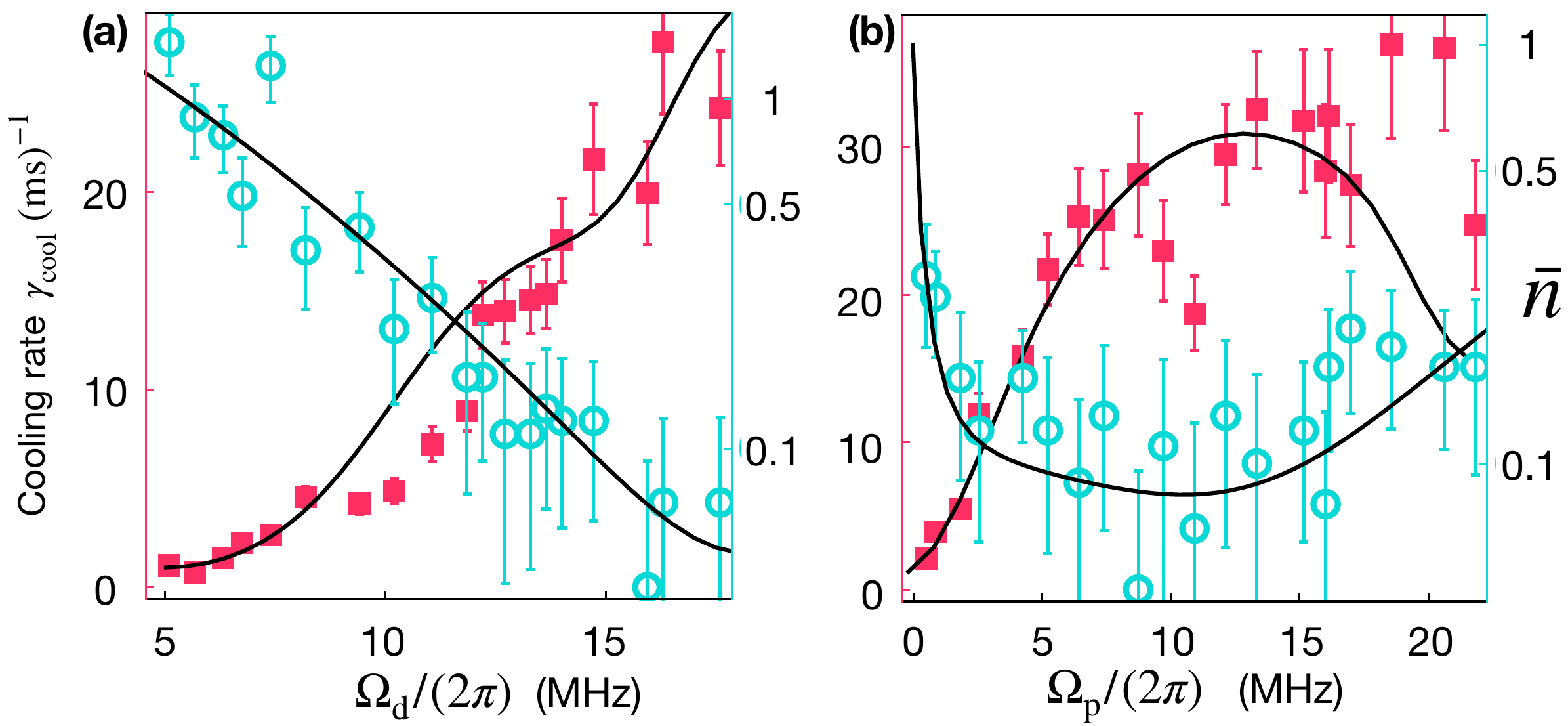}}
\caption{\label{fig:power} The final mean phonon numbers (circular points) and the cooling rates (square points) are plotted versus the power of (a) the driving beam and (b) the probe beam. Error bars denote the fitting uncertainties of blue-sideband evolutions, similar to Fig. \ref{fig:speed}(c,d). Solid lines are numerical simulation results obtained by solving the master equation \cite{supple}. }
\end{figure}	

The EIT cooling rate $\gamma_{\rm cool}$ and the minimum phonon number $n_{\min}$ as functions of intensities of the EIT beams are shown in Fig. \ref{fig:power}. We characterize the cooling efficiency as the power of the driving (probe) beam varies while the power of the probe (driving) beam is fixed at 5.5$\mu$W, $\Omega_{\rm p}/2\pi$ = 6.67 MHz (24$\mu$W, $\Omega_{\rm d}/2\pi$ = 17.39 MHz). At each point of laser powers, we search the optimal EIT detuning ($\Delta_{\rm p}-\Delta_{\rm d}$). As shown in Fig. \ref{fig:power}, numerical simulations match the experimental results fairly well, while the discrepancies of cooling rates could origin from the overall power fluctuations. As the power of the driving beam increases to the maximal possible value in our experiment, the cooling efficiency is also enhanced, as shown in Fig.\ref{fig:power}(a). On the other hand, Fig.\ref{fig:power}(b) shows that both cooling rate and limit have a local optimum. To balance the cooling rate and limit, we determine the optimal power of the probe beam by minimizing the ratio between final phonon number and cooling rate of the numerical curve in Fig. \ref{fig:power}(b). Finally, we found $\Omega_{\rm p}/(2\pi)=11$ MHz is optimal for cooling.

\begin{figure}[!htb]
\center{\includegraphics[width=0.5\textwidth]{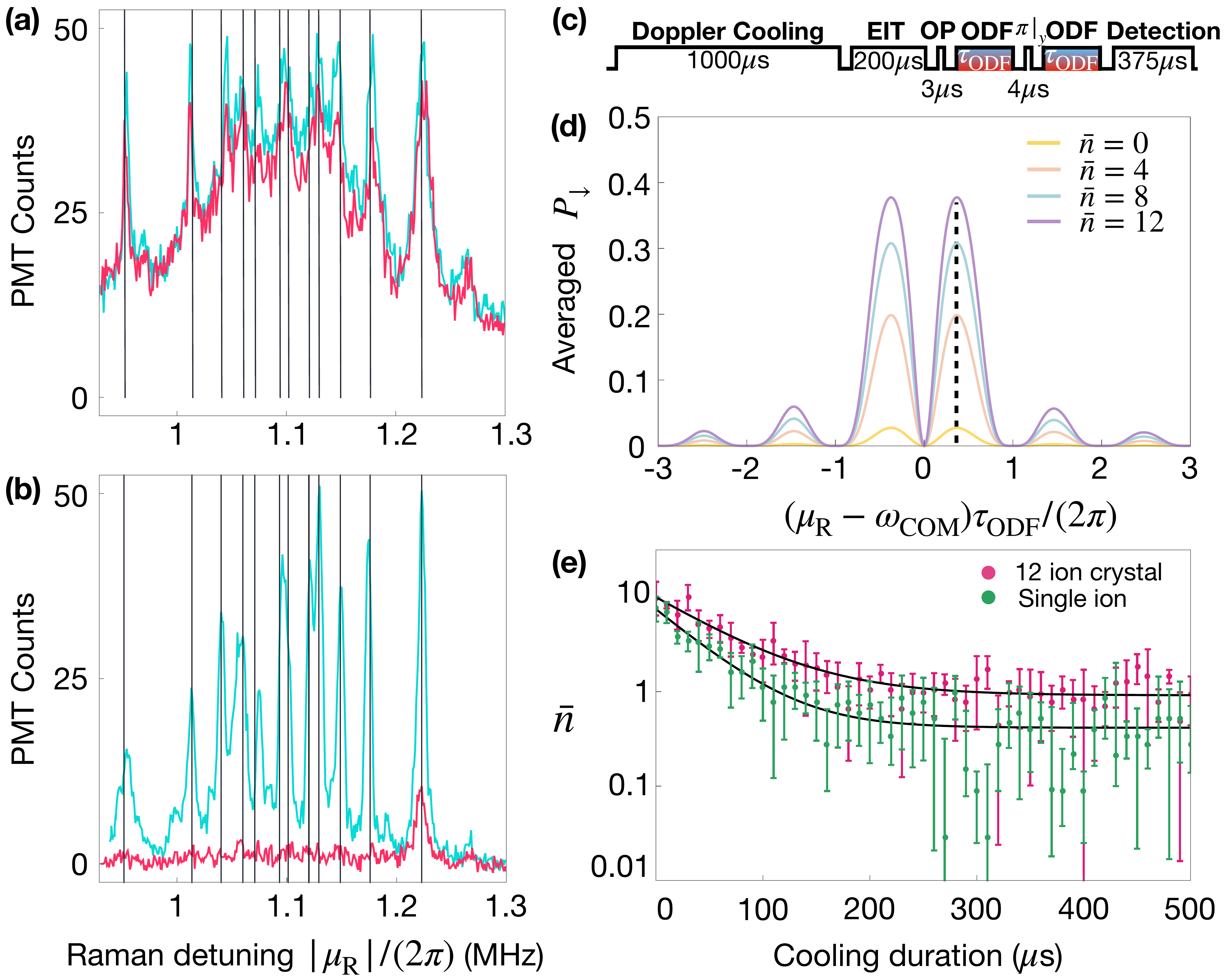}}
\caption{\label{fig:2D_crystal} (a,b) Blue-sideband (blue curve) and red-sideband (red curve) spectrum after (a) Doppler cooling and (b) EIT cooling. The vertical axis represents the count globally collected by PMT. The horizontal axis $\mu_{\rm R}=\omega_{\rm R}-\omega_{0}$ is the detuning of the Raman transition from the qubit transition. Vertical lines indicate the locations of 12 motional modes perpendicular to the 2D-crystal plane. (c) Pulse sequence for the ODF thermometry. (d) ODF spectrum with different average phonon number. The dashed black line indicates the position we choose for the cooling rate measurement. (e) Cooling dynamics for a single ion (green) and a 2D crystal with 12 ions (red). The dots are experimental data. The error bars represent the standard deviation induced by the quantum projection noise. Solid lines are fitting curves by exponential decay functions. }
\end{figure}

To assess double-EIT cooling on a large ion crystal, we store a 2D crystal of 12 ions in a pancake harmonic potentials with secular trap frequencies $\omega_{\rm y}/(2\pi)=1.22\text{MHz}$ in the transverse direction, and $\{\omega_{\rm x},\omega_{\rm z}\}/(2\pi)=\{0.34,0.42\}\text{MHz}$ in the crystal plane. With this smaller $\omega_{\rm y}$, the heating rate is increased to $0.77\rm{ms}^{-1}$. We suppress the micromotion of the 2D crystal in the transverse modes by adjusting the plane of the crystal to be in line with the micromotion direction, which is the z-axis shown in Fig. \ref{fig:setup}(c). Then, the direction of dominant micromotion is perpendicular to directions of the transverse modes and the net-propagation direction of the EIT  beams. In such a situation, the effect of micromotion is eliminated in double-EIT cooling; therefore, we can perform efficient cooling. Indeed, we measure the strength of the micromotion sideband in the Raman spectroscopy and observe it is at a similar level to a single ion \cite{wang2019realization}. Double-EIT cooling of a 12-ion crystal is observed from Raman absorption spectrum. Fig. \ref{fig:2D_crystal} (a) depicts the spectrum with only Doppler cooling, where the peaks of blue-sideband (blue curve) and red-sideband (red curve) transitions possess similar heights across all motional modes, which indicates the phonon numbers are much larger than 1. Fig.\ref{fig:2D_crystal} (b) shows the spectrum after both Doppler and EIT cooling, where the reduction of red-sideband transitions indicates simultaneous ground-state cooling of all transverse modes. The small peak in the spectrum of red-sideband transitions originates from imperfect ground-state cooling of the center of mass (COM) mode with linearly scaled heating rate. We numerically simulate the red-sideband absorption spectrum of the crystal in the vicinity of each mode for the parameters of our experiment \cite{lechner2016electromagnetically}. The estimated phonon numbers of COM mode is 1.04 $(\pm 0.26)$ \cite{supple}.  

We also use the optical-dipole-force (ODF) thermometry \cite{sawyer2012spectroscopy} to measure the final phonon number of COM mode. The ODF is generated by simultaneously driving red-sideband and blue-sideband transitions, where the $\sigma_{\rm x}\sigma_{\rm x}$ interaction emerges. With ion-phonon coupling, this $\sigma_{\rm x}\sigma_{\rm x}$ interaction could induce decoherence in the $\rm x$ basis. The Ramsey measurement is adopted to probe this decoherence, as shown in Fig. \ref{fig:2D_crystal}(c). We first prepare all qubits to its ground state in $\sigma_{\rm z}$ basis, $\ket{\downarrow}_{\rm z}$, then apply the ODF for two fixed duration $\tau_{\rm ODF}$ with a spin-echo pulse sandwiched in between. Fig. \ref{fig:2D_crystal} (d) shows the spectrum near the COM mode of the crystal with different phonon number. The temperature of the crystal is measured by fitting the spectrum to the formula [Eq. (35)] in \cite{supple}, where the $\bar{n}$ of Doppler cooling and EIT cooling are $10.72~(\pm 4.23)$ and $1.04~(\pm 0.61)$, respectively. Here, we calibrate the strength of the ODF beams by measuring the Lamb-Dicke parameter and the Rabi frequency of carrier transition \cite{supple}.

To explore the cooling dynamics for the COM mode of a crystal with 12 ions, we develop a simple method to estimate $\bar{n}$ instead of using whole ODF spectrum in Fig. \ref{fig:2D_crystal}(d). By fixing the detuning at the position with the largest decoherence, the heights of the ODF signal is converted to the mean phonon number $\bar n$ \cite{supple}. We observe a cooling rate and a cooling limit consistent with the rate and the limit of a single ion. In the setting of 12 ions, the cooling rate and the limit of a single ion are measured as $22.1~(\pm 0.1)~\rm ms^{-1}$ and $0.34 ~(\pm 0.25)$, respectively, as shown in Fig. \ref{fig:2D_crystal}(e). With 12 ions, the rate and the limit are $15.9~(\pm 0.1)~\rm ms^{-1}$ and $1.04 ~(\pm 0.61)$, respectively. The cooling rate is reduced, and the cooling limit is increased for 12 ions due to the heating rates proportional to the number of ions, which is $0.61~(\pm 0.08)~\rm ms^{-1}$ per ion \cite{supple}. In our experiment, we do not observe the more efficient EIT cooling due to the many-body effect reported in Refs.  \cite{jordan2019near,shankar2019modeling} within our error bars, which may need further experimental or theoretical study. 

In summary, we have experimentally shown that double-EIT cooling can be performed with \Yb ions and used to efficiently cool down the transverse motional modes of a 2D crystal. We demonstrate that EIT cooling can be realized for atoms and ions with more complicated level structures than $\Lambda$-scheme. Our experimental approach is suitable for other hyperfine ions with clock state, offering a fast ground-state cooling technology for atoms with long coherence time. This method may be useful for a large scale trapped ion quantum processor, for use in quantum computation, quantum magnetism, quantum chemistry, and quantum machine learning. In future work, it would be interesting to engineer the absorption spectrum for more efficient cooling \cite{scharnhorst2018experimental} and to study whether the in-plane modes of 2D crystals can be efficiently cooled down to near ground state.

\begin{acknowledgements}
We thank Paul Haljan, Dzmitry Matsukevich, Yiheng Lin, John Bollinger, Athreya Shankar, and Christian Roos for their helpful discussion. We acknowledge the use of the Quantum Toolbox in PYTHON (QuTiP) \cite{johansson2013qutip}. This work was supported by the National Key Research and Development Program of China under Grants No. 2016YFA0301900 and No. 2016YFA0301901 and the National Natural Science Foundation of China Grants No. 11374178, No. 11574002, and No. 11974200.
\end{acknowledgements}

\bibliography{arxiv}

\begin{thebibliography}{60}%
\makeatletter
\providecommand \@ifxundefined [1]{%
 \@ifx{#1\undefined}
}%
\providecommand \@ifnum [1]{%
 \ifnum #1\expandafter \@firstoftwo
 \else \expandafter \@secondoftwo
 \fi
}%
\providecommand \@ifx [1]{%
 \ifx #1\expandafter \@firstoftwo
 \else \expandafter \@secondoftwo
 \fi
}%
\providecommand \natexlab [1]{#1}%
\providecommand \enquote  [1]{``#1''}%
\providecommand \bibnamefont  [1]{#1}%
\providecommand \bibfnamefont [1]{#1}%
\providecommand \citenamefont [1]{#1}%
\providecommand \href@noop [0]{\@secondoftwo}%
\providecommand \href [0]{\begingroup \@sanitize@url \@href}%
\providecommand \@href[1]{\@@startlink{#1}\@@href}%
\providecommand \@@href[1]{\endgroup#1\@@endlink}%
\providecommand \@sanitize@url [0]{\catcode `\\12\catcode `\$12\catcode
  `\&12\catcode `\#12\catcode `\^12\catcode `\_12\catcode `\%12\relax}%
\providecommand \@@startlink[1]{}%
\providecommand \@@endlink[0]{}%
\providecommand \url  [0]{\begingroup\@sanitize@url \@url }%
\providecommand \@url [1]{\endgroup\@href {#1}{\urlprefix }}%
\providecommand \urlprefix  [0]{URL }%
\providecommand \Eprint [0]{\href }%
\providecommand \doibase [0]{http://dx.doi.org/}%
\providecommand \selectlanguage [0]{\@gobble}%
\providecommand \bibinfo  [0]{\@secondoftwo}%
\providecommand \bibfield  [0]{\@secondoftwo}%
\providecommand \translation [1]{[#1]}%
\providecommand \BibitemOpen [0]{}%
\providecommand \bibitemStop [0]{}%
\providecommand \bibitemNoStop [0]{.\EOS\space}%
\providecommand \EOS [0]{\spacefactor3000\relax}%
\providecommand \BibitemShut  [1]{\csname bibitem#1\endcsname}%
\let\auto@bib@innerbib\@empty
\bibitem [{\citenamefont {Wineland}(2013)}]{Wineland2013Novel}%
  \BibitemOpen
  \bibfield  {author} {\bibinfo {author} {\bibfnamefont {D.~J.}\ \bibnamefont
  {Wineland}},\ }\href {\doibase 10.1103/RevModPhys.85.1103} {\bibfield
  {journal} {\bibinfo  {journal} {Rev. Mod. Phys.}\ }\textbf {\bibinfo {volume}
  {85}},\ \bibinfo {pages} {1103} (\bibinfo {year} {2013})}\BibitemShut
  {NoStop}%
\bibitem [{\citenamefont {Lloyd}\ and\ \citenamefont
  {Braunstein}(1999)}]{Lloyd1999Quantum}%
  \BibitemOpen
  \bibfield  {author} {\bibinfo {author} {\bibfnamefont {S.}~\bibnamefont
  {Lloyd}}\ and\ \bibinfo {author} {\bibfnamefont {S.~L.}\ \bibnamefont
  {Braunstein}},\ }\href {\doibase 10.1103/PhysRevLett.82.1784} {\bibfield
  {journal} {\bibinfo  {journal} {Phys. Rev. Lett.}\ }\textbf {\bibinfo
  {volume} {82}},\ \bibinfo {pages} {1784} (\bibinfo {year}
  {1999})}\BibitemShut {NoStop}%
\bibitem [{\citenamefont {Lau}\ and\ \citenamefont
  {Plenio}(2016)}]{Lau2016Universal}%
  \BibitemOpen
  \bibfield  {author} {\bibinfo {author} {\bibfnamefont {H.-K.}\ \bibnamefont
  {Lau}}\ and\ \bibinfo {author} {\bibfnamefont {M.~B.}\ \bibnamefont
  {Plenio}},\ }\href {\doibase 10.1103/PhysRevLett.117.100501} {\bibfield
  {journal} {\bibinfo  {journal} {Phys. Rev. Lett.}\ }\textbf {\bibinfo
  {volume} {117}},\ \bibinfo {pages} {100501} (\bibinfo {year}
  {2016})}\BibitemShut {NoStop}%
\bibitem [{\citenamefont {Ding}\ \emph
  {et~al.}(2017{\natexlab{a}})\citenamefont {Ding}, \citenamefont
  {Maslennikov}, \citenamefont {Habl\"utzel}, \citenamefont {Loh},\ and\
  \citenamefont {Matsukevich}}]{Ding2017Quantum}%
  \BibitemOpen
  \bibfield  {author} {\bibinfo {author} {\bibfnamefont {S.}~\bibnamefont
  {Ding}}, \bibinfo {author} {\bibfnamefont {G.}~\bibnamefont {Maslennikov}},
  \bibinfo {author} {\bibfnamefont {R.}~\bibnamefont {Habl\"utzel}}, \bibinfo
  {author} {\bibfnamefont {H.}~\bibnamefont {Loh}}, \ and\ \bibinfo {author}
  {\bibfnamefont {D.}~\bibnamefont {Matsukevich}},\ }\href {\doibase
  10.1103/PhysRevLett.119.150404} {\bibfield  {journal} {\bibinfo  {journal}
  {Phys. Rev. Lett.}\ }\textbf {\bibinfo {volume} {119}},\ \bibinfo {pages}
  {150404} (\bibinfo {year} {2017}{\natexlab{a}})}\BibitemShut {NoStop}%
\bibitem [{\citenamefont {Ding}\ \emph
  {et~al.}(2017{\natexlab{b}})\citenamefont {Ding}, \citenamefont
  {Maslennikov}, \citenamefont {Habl\"utzel},\ and\ \citenamefont
  {Matsukevich}}]{Ding2017Cross}%
  \BibitemOpen
  \bibfield  {author} {\bibinfo {author} {\bibfnamefont {S.}~\bibnamefont
  {Ding}}, \bibinfo {author} {\bibfnamefont {G.}~\bibnamefont {Maslennikov}},
  \bibinfo {author} {\bibfnamefont {R.}~\bibnamefont {Habl\"utzel}}, \ and\
  \bibinfo {author} {\bibfnamefont {D.}~\bibnamefont {Matsukevich}},\ }\href
  {\doibase 10.1103/PhysRevLett.119.193602} {\bibfield  {journal} {\bibinfo
  {journal} {Phys. Rev. Lett.}\ }\textbf {\bibinfo {volume} {119}},\ \bibinfo
  {pages} {193602} (\bibinfo {year} {2017}{\natexlab{b}})}\BibitemShut
  {NoStop}%
\bibitem [{\citenamefont {Fl{\"u}hmann}\ \emph {et~al.}(2019)\citenamefont
  {Fl{\"u}hmann}, \citenamefont {Nguyen}, \citenamefont {Marinelli},
  \citenamefont {Negnevitsky}, \citenamefont {Mehta},\ and\ \citenamefont
  {Home}}]{fluhmann2019encoding}%
  \BibitemOpen
  \bibfield  {author} {\bibinfo {author} {\bibfnamefont {C.}~\bibnamefont
  {Fl{\"u}hmann}}, \bibinfo {author} {\bibfnamefont {T.~L.}\ \bibnamefont
  {Nguyen}}, \bibinfo {author} {\bibfnamefont {M.}~\bibnamefont {Marinelli}},
  \bibinfo {author} {\bibfnamefont {V.}~\bibnamefont {Negnevitsky}}, \bibinfo
  {author} {\bibfnamefont {K.}~\bibnamefont {Mehta}}, \ and\ \bibinfo {author}
  {\bibfnamefont {J.}~\bibnamefont {Home}},\ }\href
  {https://www.nature.com/articles/s41586-019-0960-6} {\bibfield  {journal}
  {\bibinfo  {journal} {Nature}\ }\textbf {\bibinfo {volume} {566}},\ \bibinfo
  {pages} {513} (\bibinfo {year} {2019})}\BibitemShut {NoStop}%
\bibitem [{\citenamefont {Aaronson}\ and\ \citenamefont
  {Arkhipov}(2011)}]{Aaronson2011}%
  \BibitemOpen
  \bibfield  {author} {\bibinfo {author} {\bibfnamefont {S.}~\bibnamefont
  {Aaronson}}\ and\ \bibinfo {author} {\bibfnamefont {A.}~\bibnamefont
  {Arkhipov}},\ }\href {\doibase 10.1145/1993636.1993682} {\bibfield  {journal}
  {\bibinfo  {journal} {Proceedings of the 43rd annual ACM symposium on Theory
  of computing - STOC '11}\ ,\ \bibinfo {pages} {333}} (\bibinfo {year}
  {2011})}\BibitemShut {NoStop}%
\bibitem [{\citenamefont {Lau}\ and\ \citenamefont
  {James}(2012)}]{Lau2012Proposal}%
  \BibitemOpen
  \bibfield  {author} {\bibinfo {author} {\bibfnamefont {H.-K.}\ \bibnamefont
  {Lau}}\ and\ \bibinfo {author} {\bibfnamefont {D.~F.~V.}\ \bibnamefont
  {James}},\ }\href {\doibase 10.1103/PhysRevA.85.062329} {\bibfield  {journal}
  {\bibinfo  {journal} {Phys. Rev. A}\ }\textbf {\bibinfo {volume} {85}},\
  \bibinfo {pages} {062329} (\bibinfo {year} {2012})}\BibitemShut {NoStop}%
\bibitem [{\citenamefont {Shen}\ \emph {et~al.}(2014)\citenamefont {Shen},
  \citenamefont {Zhang},\ and\ \citenamefont {Duan}}]{shen2014scalable}%
  \BibitemOpen
  \bibfield  {author} {\bibinfo {author} {\bibfnamefont {C.}~\bibnamefont
  {Shen}}, \bibinfo {author} {\bibfnamefont {Z.}~\bibnamefont {Zhang}}, \ and\
  \bibinfo {author} {\bibfnamefont {L.-M.}\ \bibnamefont {Duan}},\ }\href
  {https://link.aps.org/doi/10.1103/PhysRevLett.112.050504} {\bibfield
  {journal} {\bibinfo  {journal} {Phys. Rev. Lett.}\ }\textbf {\bibinfo
  {volume} {112}},\ \bibinfo {pages} {050504} (\bibinfo {year}
  {2014})}\BibitemShut {NoStop}%
\bibitem [{\citenamefont {Toyoda}\ \emph {et~al.}(2015)\citenamefont {Toyoda},
  \citenamefont {Hiji}, \citenamefont {Noguchi},\ and\ \citenamefont
  {Urabe}}]{toyoda2015hon}%
  \BibitemOpen
  \bibfield  {author} {\bibinfo {author} {\bibfnamefont {K.}~\bibnamefont
  {Toyoda}}, \bibinfo {author} {\bibfnamefont {R.}~\bibnamefont {Hiji}},
  \bibinfo {author} {\bibfnamefont {A.}~\bibnamefont {Noguchi}}, \ and\
  \bibinfo {author} {\bibfnamefont {S.}~\bibnamefont {Urabe}},\ }\href
  {https://www.nature.com/articles/ncomms11410} {\bibfield  {journal} {\bibinfo
   {journal} {Nature}\ }\textbf {\bibinfo {volume} {527}},\ \bibinfo {pages}
  {74} (\bibinfo {year} {2015})}\BibitemShut {NoStop}%
\bibitem [{\citenamefont {Um}\ \emph {et~al.}(2016)\citenamefont {Um},
  \citenamefont {Zhang}, \citenamefont {Lv}, \citenamefont {Lu}, \citenamefont
  {An}, \citenamefont {Zhang}, \citenamefont {Nha}, \citenamefont {Kim},\ and\
  \citenamefont {Kim}}]{um2016phonon}%
  \BibitemOpen
  \bibfield  {author} {\bibinfo {author} {\bibfnamefont {M.}~\bibnamefont
  {Um}}, \bibinfo {author} {\bibfnamefont {J.}~\bibnamefont {Zhang}}, \bibinfo
  {author} {\bibfnamefont {D.}~\bibnamefont {Lv}}, \bibinfo {author}
  {\bibfnamefont {Y.}~\bibnamefont {Lu}}, \bibinfo {author} {\bibfnamefont
  {S.}~\bibnamefont {An}}, \bibinfo {author} {\bibfnamefont {J.-N.}\
  \bibnamefont {Zhang}}, \bibinfo {author} {\bibfnamefont {H.}~\bibnamefont
  {Nha}}, \bibinfo {author} {\bibfnamefont {M.}~\bibnamefont {Kim}}, \ and\
  \bibinfo {author} {\bibfnamefont {K.}~\bibnamefont {Kim}},\ }\href@noop {}
  {\bibfield  {journal} {\bibinfo  {journal} {Nat. Commun.}\ }\textbf {\bibinfo
  {volume} {7}},\ \bibinfo {pages} {1} (\bibinfo {year} {2016})}\BibitemShut
  {NoStop}%
\bibitem [{\citenamefont {Shen}\ \emph {et~al.}(2018)\citenamefont {Shen},
  \citenamefont {Lu}, \citenamefont {Zhang}, \citenamefont {Zhang},
  \citenamefont {Zhang}, \citenamefont {Huh},\ and\ \citenamefont
  {Kim}}]{shen2018quantum}%
  \BibitemOpen
  \bibfield  {author} {\bibinfo {author} {\bibfnamefont {Y.}~\bibnamefont
  {Shen}}, \bibinfo {author} {\bibfnamefont {Y.}~\bibnamefont {Lu}}, \bibinfo
  {author} {\bibfnamefont {K.}~\bibnamefont {Zhang}}, \bibinfo {author}
  {\bibfnamefont {J.}~\bibnamefont {Zhang}}, \bibinfo {author} {\bibfnamefont
  {S.}~\bibnamefont {Zhang}}, \bibinfo {author} {\bibfnamefont
  {J.}~\bibnamefont {Huh}}, \ and\ \bibinfo {author} {\bibfnamefont
  {K.}~\bibnamefont {Kim}},\ }\href
  {https://pubs.rsc.org/en/content/articlelanding/2018/sc/c7sc04602b#!divAbstract}
  {\bibfield  {journal} {\bibinfo  {journal} {Chem. Sci.}\ }\textbf {\bibinfo
  {volume} {9}},\ \bibinfo {pages} {836} (\bibinfo {year} {2018})}\BibitemShut
  {NoStop}%
\bibitem [{\citenamefont {Ballance}\ \emph {et~al.}(2016)\citenamefont
  {Ballance}, \citenamefont {Harty}, \citenamefont {Linke}, \citenamefont
  {Sepiol},\ and\ \citenamefont {Lucas}}]{ballance2016high}%
  \BibitemOpen
  \bibfield  {author} {\bibinfo {author} {\bibfnamefont {C.~J.}\ \bibnamefont
  {Ballance}}, \bibinfo {author} {\bibfnamefont {T.~P.}\ \bibnamefont {Harty}},
  \bibinfo {author} {\bibfnamefont {N.~M.}\ \bibnamefont {Linke}}, \bibinfo
  {author} {\bibfnamefont {M.~A.}\ \bibnamefont {Sepiol}}, \ and\ \bibinfo
  {author} {\bibfnamefont {D.~M.}\ \bibnamefont {Lucas}},\ }\href
  {https://journals.aps.org/prl/abstract/10.1103/PhysRevLett.117.060504}
  {\bibfield  {journal} {\bibinfo  {journal} {Phys. Rev. Lett.}\ }\textbf
  {\bibinfo {volume} {117}},\ \bibinfo {pages} {060504} (\bibinfo {year}
  {2016})}\BibitemShut {NoStop}%
\bibitem [{\citenamefont {Harty}\ \emph {et~al.}(2016)\citenamefont {Harty},
  \citenamefont {Sepiol}, \citenamefont {Allcock}, \citenamefont {Ballance},
  \citenamefont {Tarlton},\ and\ \citenamefont {Lucas}}]{harty2016high}%
  \BibitemOpen
  \bibfield  {author} {\bibinfo {author} {\bibfnamefont {T.~P.}\ \bibnamefont
  {Harty}}, \bibinfo {author} {\bibfnamefont {M.~A.}\ \bibnamefont {Sepiol}},
  \bibinfo {author} {\bibfnamefont {D.~T.~C.}\ \bibnamefont {Allcock}},
  \bibinfo {author} {\bibfnamefont {C.~J.}\ \bibnamefont {Ballance}}, \bibinfo
  {author} {\bibfnamefont {J.~E.}\ \bibnamefont {Tarlton}}, \ and\ \bibinfo
  {author} {\bibfnamefont {D.~M.}\ \bibnamefont {Lucas}},\ }\href
  {https://link.aps.org/pdf/10.1103/PhysRevLett.117.140501} {\bibfield
  {journal} {\bibinfo  {journal} {Phys. Rev. Lett.}\ }\textbf {\bibinfo
  {volume} {117}},\ \bibinfo {pages} {140501} (\bibinfo {year}
  {2016})}\BibitemShut {NoStop}%
\bibitem [{\citenamefont {Blatt}\ and\ \citenamefont
  {Roos}(2012)}]{blatt2012quantum}%
  \BibitemOpen
  \bibfield  {author} {\bibinfo {author} {\bibfnamefont {R.}~\bibnamefont
  {Blatt}}\ and\ \bibinfo {author} {\bibfnamefont {C.~F.}\ \bibnamefont
  {Roos}},\ }\href {https://www.nature.com/articles/nphys2252/} {\bibfield
  {journal} {\bibinfo  {journal} {Nature Physics}\ }\textbf {\bibinfo {volume}
  {8}},\ \bibinfo {pages} {277} (\bibinfo {year} {2012})}\BibitemShut {NoStop}%
\bibitem [{\citenamefont {Monroe}\ \emph {et~al.}(2019)\citenamefont {Monroe},
  \citenamefont {Campbell}, \citenamefont {Duan}, \citenamefont {Gong},
  \citenamefont {Gorshkov}, \citenamefont {Hess}, \citenamefont {Islam},
  \citenamefont {Kim}, \citenamefont {Pagano}, \citenamefont {Richerme} \emph
  {et~al.}}]{monroe2019programmable}%
  \BibitemOpen
  \bibfield  {author} {\bibinfo {author} {\bibfnamefont {C.}~\bibnamefont
  {Monroe}}, \bibinfo {author} {\bibfnamefont {W.}~\bibnamefont {Campbell}},
  \bibinfo {author} {\bibfnamefont {L.-M.}\ \bibnamefont {Duan}}, \bibinfo
  {author} {\bibfnamefont {Z.-X.}\ \bibnamefont {Gong}}, \bibinfo {author}
  {\bibfnamefont {A.}~\bibnamefont {Gorshkov}}, \bibinfo {author}
  {\bibfnamefont {P.}~\bibnamefont {Hess}}, \bibinfo {author} {\bibfnamefont
  {R.}~\bibnamefont {Islam}}, \bibinfo {author} {\bibfnamefont
  {K.}~\bibnamefont {Kim}}, \bibinfo {author} {\bibfnamefont {G.}~\bibnamefont
  {Pagano}}, \bibinfo {author} {\bibfnamefont {P.}~\bibnamefont {Richerme}},
  \emph {et~al.},\ }\href {https://arxiv.org/abs/1912.07845} {\bibfield
  {journal} {\bibinfo  {journal} {arXiv preprint arXiv:1912.07845}\ } (\bibinfo
  {year} {2019})}\BibitemShut {NoStop}%
\bibitem [{\citenamefont {Mezzacapo}\ \emph {et~al.}(2012)\citenamefont
  {Mezzacapo}, \citenamefont {Casanova}, \citenamefont {Lamata},\ and\
  \citenamefont {Solano}}]{mezzacapo2012digital}%
  \BibitemOpen
  \bibfield  {author} {\bibinfo {author} {\bibfnamefont {A.}~\bibnamefont
  {Mezzacapo}}, \bibinfo {author} {\bibfnamefont {J.}~\bibnamefont {Casanova}},
  \bibinfo {author} {\bibfnamefont {L.}~\bibnamefont {Lamata}}, \ and\ \bibinfo
  {author} {\bibfnamefont {E.}~\bibnamefont {Solano}},\ }\href
  {https://journals.aps.org/prl/abstract/10.1103/PhysRevLett.109.200501}
  {\bibfield  {journal} {\bibinfo  {journal} {Phys. Rev. Lett.}\ }\textbf
  {\bibinfo {volume} {109}},\ \bibinfo {pages} {200501} (\bibinfo {year}
  {2012})}\BibitemShut {NoStop}%
\bibitem [{\citenamefont {Lv}\ \emph {et~al.}(2018)\citenamefont {Lv},
  \citenamefont {An}, \citenamefont {Liu}, \citenamefont {Zhang}, \citenamefont
  {Pedernales}, \citenamefont {Lamata}, \citenamefont {Solano},\ and\
  \citenamefont {Kim}}]{lv2018quantum}%
  \BibitemOpen
  \bibfield  {author} {\bibinfo {author} {\bibfnamefont {D.}~\bibnamefont
  {Lv}}, \bibinfo {author} {\bibfnamefont {S.}~\bibnamefont {An}}, \bibinfo
  {author} {\bibfnamefont {Z.}~\bibnamefont {Liu}}, \bibinfo {author}
  {\bibfnamefont {J.-N.}\ \bibnamefont {Zhang}}, \bibinfo {author}
  {\bibfnamefont {J.~S.}\ \bibnamefont {Pedernales}}, \bibinfo {author}
  {\bibfnamefont {L.}~\bibnamefont {Lamata}}, \bibinfo {author} {\bibfnamefont
  {E.}~\bibnamefont {Solano}}, \ and\ \bibinfo {author} {\bibfnamefont
  {K.}~\bibnamefont {Kim}},\ }\href
  {https://journals.aps.org/prx/abstract/10.1103/PhysRevX.8.021027} {\bibfield
  {journal} {\bibinfo  {journal} {Physical Review X}\ }\textbf {\bibinfo
  {volume} {8}},\ \bibinfo {pages} {021027} (\bibinfo {year}
  {2018})}\BibitemShut {NoStop}%
\bibitem [{\citenamefont {Wineland}\ \emph {et~al.}(1978)\citenamefont
  {Wineland}, \citenamefont {Drullinger},\ and\ \citenamefont
  {Walls}}]{wineland1978radiation}%
  \BibitemOpen
  \bibfield  {author} {\bibinfo {author} {\bibfnamefont {D.~J.}\ \bibnamefont
  {Wineland}}, \bibinfo {author} {\bibfnamefont {R.~E.}\ \bibnamefont
  {Drullinger}}, \ and\ \bibinfo {author} {\bibfnamefont {F.~L.}\ \bibnamefont
  {Walls}},\ }\href {https://link.aps.org/doi/10.1103/PhysRevLett.40.1639}
  {\bibfield  {journal} {\bibinfo  {journal} {Phys. Rev. Lett.}\ }\textbf
  {\bibinfo {volume} {40}},\ \bibinfo {pages} {1639} (\bibinfo {year}
  {1978})}\BibitemShut {NoStop}%
\bibitem [{\citenamefont {Neuhauser}\ \emph {et~al.}(1978)\citenamefont
  {Neuhauser}, \citenamefont {Hohenstatt}, \citenamefont {Toschek},\ and\
  \citenamefont {Dehmelt}}]{neuhauser1978optical}%
  \BibitemOpen
  \bibfield  {author} {\bibinfo {author} {\bibfnamefont {W.}~\bibnamefont
  {Neuhauser}}, \bibinfo {author} {\bibfnamefont {M.}~\bibnamefont
  {Hohenstatt}}, \bibinfo {author} {\bibfnamefont {P.}~\bibnamefont {Toschek}},
  \ and\ \bibinfo {author} {\bibfnamefont {H.}~\bibnamefont {Dehmelt}},\ }\href
  {https://journals.aps.org/prl/abstract/10.1103/PhysRevLett.41.233} {\bibfield
   {journal} {\bibinfo  {journal} {Phys. Rev. Lett.}\ }\textbf {\bibinfo
  {volume} {41}},\ \bibinfo {pages} {233} (\bibinfo {year} {1978})}\BibitemShut
  {NoStop}%
\bibitem [{\citenamefont {Dalibard}\ and\ \citenamefont
  {Cohen-Tannoudji}(1989)}]{dalibard1989laser}%
  \BibitemOpen
  \bibfield  {author} {\bibinfo {author} {\bibfnamefont {J.}~\bibnamefont
  {Dalibard}}\ and\ \bibinfo {author} {\bibfnamefont {C.}~\bibnamefont
  {Cohen-Tannoudji}},\ }\href
  {https://www.osapublishing.org/josab/abstract.cfm?URI=josab-6-11-2023}
  {\bibfield  {journal} {\bibinfo  {journal} {JOSA B}\ }\textbf {\bibinfo
  {volume} {6}},\ \bibinfo {pages} {2023} (\bibinfo {year} {1989})}\BibitemShut
  {NoStop}%
\bibitem [{\citenamefont {Ejtemaee}\ and\ \citenamefont
  {Haljan}(2017)}]{ejtemaee20173d}%
  \BibitemOpen
  \bibfield  {author} {\bibinfo {author} {\bibfnamefont {S.}~\bibnamefont
  {Ejtemaee}}\ and\ \bibinfo {author} {\bibfnamefont {P.~C.}\ \bibnamefont
  {Haljan}},\ }\href
  {https://journals.aps.org/prl/abstract/10.1103/PhysRevLett.119.043001}
  {\bibfield  {journal} {\bibinfo  {journal} {Phys. Rev. Lett.}\ }\textbf
  {\bibinfo {volume} {119}},\ \bibinfo {pages} {043001} (\bibinfo {year}
  {2017})}\BibitemShut {NoStop}%
\bibitem [{\citenamefont {Monroe}\ \emph {et~al.}(1995)\citenamefont {Monroe},
  \citenamefont {Meekhof}, \citenamefont {King}, \citenamefont {Jefferts},
  \citenamefont {Itano}, \citenamefont {Wineland},\ and\ \citenamefont
  {Gould}}]{monroe1995resolved}%
  \BibitemOpen
  \bibfield  {author} {\bibinfo {author} {\bibfnamefont {C.}~\bibnamefont
  {Monroe}}, \bibinfo {author} {\bibfnamefont {D.~M.}\ \bibnamefont {Meekhof}},
  \bibinfo {author} {\bibfnamefont {B.~E.}\ \bibnamefont {King}}, \bibinfo
  {author} {\bibfnamefont {S.~R.}\ \bibnamefont {Jefferts}}, \bibinfo {author}
  {\bibfnamefont {W.~M.}\ \bibnamefont {Itano}}, \bibinfo {author}
  {\bibfnamefont {D.~J.}\ \bibnamefont {Wineland}}, \ and\ \bibinfo {author}
  {\bibfnamefont {P.}~\bibnamefont {Gould}},\ }\href
  {https://journals.aps.org/prl/abstract/10.1103/PhysRevLett.75.4011}
  {\bibfield  {journal} {\bibinfo  {journal} {Phys. Rev. Lett.}\ }\textbf
  {\bibinfo {volume} {75}},\ \bibinfo {pages} {4011} (\bibinfo {year}
  {1995})}\BibitemShut {NoStop}%
\bibitem [{\citenamefont {Roos}\ \emph {et~al.}(1999)\citenamefont {Roos},
  \citenamefont {Zeiger}, \citenamefont {Rohde}, \citenamefont {N{\"a}gerl},
  \citenamefont {Eschner}, \citenamefont {Leibfried}, \citenamefont
  {Schmidt-Kaler},\ and\ \citenamefont {Blatt}}]{roos1999quantum}%
  \BibitemOpen
  \bibfield  {author} {\bibinfo {author} {\bibfnamefont {C.}~\bibnamefont
  {Roos}}, \bibinfo {author} {\bibfnamefont {T.}~\bibnamefont {Zeiger}},
  \bibinfo {author} {\bibfnamefont {H.}~\bibnamefont {Rohde}}, \bibinfo
  {author} {\bibfnamefont {H.~C.}\ \bibnamefont {N{\"a}gerl}}, \bibinfo
  {author} {\bibfnamefont {J.}~\bibnamefont {Eschner}}, \bibinfo {author}
  {\bibfnamefont {D.}~\bibnamefont {Leibfried}}, \bibinfo {author}
  {\bibfnamefont {F.}~\bibnamefont {Schmidt-Kaler}}, \ and\ \bibinfo {author}
  {\bibfnamefont {R.}~\bibnamefont {Blatt}},\ }\href
  {https://journals.aps.org/prl/abstract/10.1103/PhysRevLett.83.4713}
  {\bibfield  {journal} {\bibinfo  {journal} {Phys. Rev. Lett.}\ }\textbf
  {\bibinfo {volume} {83}},\ \bibinfo {pages} {4713} (\bibinfo {year}
  {1999})}\BibitemShut {NoStop}%
\bibitem [{\citenamefont {Harlander}\ \emph {et~al.}(2010)\citenamefont
  {Harlander}, \citenamefont {Brownnutt}, \citenamefont {H{\"a}nsel},\ and\
  \citenamefont {Blatt}}]{harlander2010trapped}%
  \BibitemOpen
  \bibfield  {author} {\bibinfo {author} {\bibfnamefont {M.}~\bibnamefont
  {Harlander}}, \bibinfo {author} {\bibfnamefont {M.}~\bibnamefont
  {Brownnutt}}, \bibinfo {author} {\bibfnamefont {W.}~\bibnamefont
  {H{\"a}nsel}}, \ and\ \bibinfo {author} {\bibfnamefont {R.}~\bibnamefont
  {Blatt}},\ }\href
  {https://iopscience.iop.org/article/10.1088/1367-2630/12/9/093035} {\bibfield
   {journal} {\bibinfo  {journal} {New Journal of Physics}\ }\textbf {\bibinfo
  {volume} {12}},\ \bibinfo {pages} {093035} (\bibinfo {year}
  {2010})}\BibitemShut {NoStop}%
\bibitem [{sup()}]{supple}%
  \BibitemOpen
  \href@noop {} {\bibinfo  {journal} {See Supplemental Material [url] for more
  details on the theory of double-EIT cooling, the calibration of AC stark shift,
  and the measurement of phonon number, which includes Refs. [45-47, 50-52]}\ }\BibitemShut
  {NoStop}%
\bibitem [{\citenamefont {Morigi}\ \emph {et~al.}(2000)\citenamefont {Morigi},
  \citenamefont {Eschner},\ and\ \citenamefont {Keitel}}]{morigi2000ground}%
  \BibitemOpen
\bibfield  {journal} {  }\bibfield  {author} {\bibinfo {author} {\bibfnamefont
  {G.}~\bibnamefont {Morigi}}, \bibinfo {author} {\bibfnamefont
  {J.}~\bibnamefont {Eschner}}, \ and\ \bibinfo {author} {\bibfnamefont
  {C.~H.}\ \bibnamefont {Keitel}},\ }\href
  {https://journals.aps.org/prl/abstract/10.1103/PhysRevLett.85.4458}
  {\bibfield  {journal} {\bibinfo  {journal} {Phys. Rev. Lett.}\ }\textbf
  {\bibinfo {volume} {85}},\ \bibinfo {pages} {4458} (\bibinfo {year}
  {2000})}\BibitemShut {NoStop}%
\bibitem [{\citenamefont {Roos}\ \emph {et~al.}(2000)\citenamefont {Roos},
  \citenamefont {Leibfried}, \citenamefont {Mundt}, \citenamefont
  {Schmidt-Kaler}, \citenamefont {Eschner},\ and\ \citenamefont
  {Blatt}}]{roos2000experimental}%
  \BibitemOpen
  \bibfield  {author} {\bibinfo {author} {\bibfnamefont {C.~F.}\ \bibnamefont
  {Roos}}, \bibinfo {author} {\bibfnamefont {D.}~\bibnamefont {Leibfried}},
  \bibinfo {author} {\bibfnamefont {A.}~\bibnamefont {Mundt}}, \bibinfo
  {author} {\bibfnamefont {F.}~\bibnamefont {Schmidt-Kaler}}, \bibinfo {author}
  {\bibfnamefont {J.}~\bibnamefont {Eschner}}, \ and\ \bibinfo {author}
  {\bibfnamefont {R.}~\bibnamefont {Blatt}},\ }\href
  {https://journals.aps.org/prl/abstract/10.1103/PhysRevLett.85.5547}
  {\bibfield  {journal} {\bibinfo  {journal} {Phys. Rev. Lett.}\ }\textbf
  {\bibinfo {volume} {85}},\ \bibinfo {pages} {5547} (\bibinfo {year}
  {2000})}\BibitemShut {NoStop}%
\bibitem [{\citenamefont {Lin}\ \emph {et~al.}(2013)\citenamefont {Lin},
  \citenamefont {Gaebler}, \citenamefont {Tan}, \citenamefont {Bowler},
  \citenamefont {Jost}, \citenamefont {Leibfried},\ and\ \citenamefont
  {Wineland}}]{lin2013sympathetic}%
  \BibitemOpen
  \bibfield  {author} {\bibinfo {author} {\bibfnamefont {Y.}~\bibnamefont
  {Lin}}, \bibinfo {author} {\bibfnamefont {J.~P.}\ \bibnamefont {Gaebler}},
  \bibinfo {author} {\bibfnamefont {T.~R.}\ \bibnamefont {Tan}}, \bibinfo
  {author} {\bibfnamefont {R.}~\bibnamefont {Bowler}}, \bibinfo {author}
  {\bibfnamefont {J.~D.}\ \bibnamefont {Jost}}, \bibinfo {author}
  {\bibfnamefont {D.}~\bibnamefont {Leibfried}}, \ and\ \bibinfo {author}
  {\bibfnamefont {D.~J.}\ \bibnamefont {Wineland}},\ }\href
  {https://journals.aps.org/prl/abstract/10.1103/PhysRevLett.110.153002}
  {\bibfield  {journal} {\bibinfo  {journal} {Phys. Rev. Lett.}\ }\textbf
  {\bibinfo {volume} {110}},\ \bibinfo {pages} {153002} (\bibinfo {year}
  {2013})}\BibitemShut {NoStop}%
\bibitem [{\citenamefont {Lechner}\ \emph {et~al.}(2016)\citenamefont
  {Lechner}, \citenamefont {Maier}, \citenamefont {Hempel}, \citenamefont
  {Jurcevic}, \citenamefont {Lanyon}, \citenamefont {Monz}, \citenamefont
  {Brownnutt}, \citenamefont {Blatt},\ and\ \citenamefont
  {Roos}}]{lechner2016electromagnetically}%
  \BibitemOpen
  \bibfield  {author} {\bibinfo {author} {\bibfnamefont {R.}~\bibnamefont
  {Lechner}}, \bibinfo {author} {\bibfnamefont {C.}~\bibnamefont {Maier}},
  \bibinfo {author} {\bibfnamefont {C.}~\bibnamefont {Hempel}}, \bibinfo
  {author} {\bibfnamefont {P.}~\bibnamefont {Jurcevic}}, \bibinfo {author}
  {\bibfnamefont {B.~P.}\ \bibnamefont {Lanyon}}, \bibinfo {author}
  {\bibfnamefont {T.}~\bibnamefont {Monz}}, \bibinfo {author} {\bibfnamefont
  {M.}~\bibnamefont {Brownnutt}}, \bibinfo {author} {\bibfnamefont
  {R.}~\bibnamefont {Blatt}}, \ and\ \bibinfo {author} {\bibfnamefont {C.~F.}\
  \bibnamefont {Roos}},\ }\href
  {https://link.aps.org/pdf/10.1103/PhysRevA.93.053401} {\bibfield  {journal}
  {\bibinfo  {journal} {Physical Review A}\ }\textbf {\bibinfo {volume} {93}},\
  \bibinfo {pages} {053401} (\bibinfo {year} {2016})}\BibitemShut {NoStop}%
\bibitem [{\citenamefont {Jordan}\ \emph {et~al.}(2019)\citenamefont {Jordan},
  \citenamefont {Gilmore}, \citenamefont {Shankar}, \citenamefont
  {Safavi-Naini}, \citenamefont {Bohnet}, \citenamefont {Holland},\ and\
  \citenamefont {Bollinger}}]{jordan2019near}%
  \BibitemOpen
  \bibfield  {author} {\bibinfo {author} {\bibfnamefont {E.}~\bibnamefont
  {Jordan}}, \bibinfo {author} {\bibfnamefont {K.~A.}\ \bibnamefont {Gilmore}},
  \bibinfo {author} {\bibfnamefont {A.}~\bibnamefont {Shankar}}, \bibinfo
  {author} {\bibfnamefont {A.}~\bibnamefont {Safavi-Naini}}, \bibinfo {author}
  {\bibfnamefont {J.~G.}\ \bibnamefont {Bohnet}}, \bibinfo {author}
  {\bibfnamefont {M.~J.}\ \bibnamefont {Holland}}, \ and\ \bibinfo {author}
  {\bibfnamefont {J.~J.}\ \bibnamefont {Bollinger}},\ }\href
  {https://journals.aps.org/prl/abstract/10.1103/PhysRevLett.122.053603}
  {\bibfield  {journal} {\bibinfo  {journal} {Phys. Rev. Lett.}\ }\textbf
  {\bibinfo {volume} {122}},\ \bibinfo {pages} {053603} (\bibinfo {year}
  {2019})}\BibitemShut {NoStop}%
\bibitem [{\citenamefont {Paspalakis}\ and\ \citenamefont
  {Knight}(2002)}]{paspalakis2002transparency}%
  \BibitemOpen
  \bibfield  {author} {\bibinfo {author} {\bibfnamefont {E.}~\bibnamefont
  {Paspalakis}}\ and\ \bibinfo {author} {\bibfnamefont {P.}~\bibnamefont
  {Knight}},\ }\href
  {https://iopscience.iop.org/article/10.1088/1464-4266/4/4/322} {\bibfield
  {journal} {\bibinfo  {journal} {Journal of Optics B: Quantum and
  Semiclassical Optics}\ }\textbf {\bibinfo {volume} {4}},\ \bibinfo {pages}
  {S372} (\bibinfo {year} {2002})}\BibitemShut {NoStop}%
\bibitem [{\citenamefont {Beck}\ and\ \citenamefont
  {Mazets}(2017)}]{beck2017propagation}%
  \BibitemOpen
  \bibfield  {author} {\bibinfo {author} {\bibfnamefont {S.}~\bibnamefont
  {Beck}}\ and\ \bibinfo {author} {\bibfnamefont {I.~E.}\ \bibnamefont
  {Mazets}},\ }\href
  {https://journals.aps.org/pra/abstract/10.1103/PhysRevA.95.013818} {\bibfield
   {journal} {\bibinfo  {journal} {Physical Review A}\ }\textbf {\bibinfo
  {volume} {95}},\ \bibinfo {pages} {013818} (\bibinfo {year}
  {2017})}\BibitemShut {NoStop}%
\bibitem [{\citenamefont {Alotaibi}\ and\ \citenamefont
  {Sanders}(2014)}]{alotaibi2014double}%
  \BibitemOpen
  \bibfield  {author} {\bibinfo {author} {\bibfnamefont {H.~M.~M.}\
  \bibnamefont {Alotaibi}}\ and\ \bibinfo {author} {\bibfnamefont {B.~C.}\
  \bibnamefont {Sanders}},\ }\href
  {https://journals.aps.org/pra/abstract/10.1103/PhysRevA.89.021802} {\bibfield
   {journal} {\bibinfo  {journal} {Physical Review A}\ }\textbf {\bibinfo
  {volume} {89}},\ \bibinfo {pages} {021802(R)} (\bibinfo {year}
  {2014})}\BibitemShut {NoStop}%
\bibitem [{\citenamefont {Ham}\ and\ \citenamefont
  {Hemmer}(2000)}]{ham2000coherence}%
  \BibitemOpen
  \bibfield  {author} {\bibinfo {author} {\bibfnamefont {B.~S.}\ \bibnamefont
  {Ham}}\ and\ \bibinfo {author} {\bibfnamefont {P.~R.}\ \bibnamefont
  {Hemmer}},\ }\href
  {https://journals.aps.org/prl/abstract/10.1103/PhysRevLett.84.4080}
  {\bibfield  {journal} {\bibinfo  {journal} {Phys. Rev. Lett.}\ }\textbf
  {\bibinfo {volume} {84}},\ \bibinfo {pages} {4080} (\bibinfo {year}
  {2000})}\BibitemShut {NoStop}%
\bibitem [{\citenamefont {Lee}\ \emph {et~al.}(2014)\citenamefont {Lee},
  \citenamefont {Ruseckas}, \citenamefont {Lee}, \citenamefont
  {Kudria{\v{s}}ov}, \citenamefont {Chang}, \citenamefont {Cho}, \citenamefont
  {Juzeli{\=a}nas},\ and\ \citenamefont {Ite}}]{lee2014experimental}%
  \BibitemOpen
  \bibfield  {author} {\bibinfo {author} {\bibfnamefont {M.-J.}\ \bibnamefont
  {Lee}}, \bibinfo {author} {\bibfnamefont {J.}~\bibnamefont {Ruseckas}},
  \bibinfo {author} {\bibfnamefont {C.-Y.}\ \bibnamefont {Lee}}, \bibinfo
  {author} {\bibfnamefont {V.}~\bibnamefont {Kudria{\v{s}}ov}}, \bibinfo
  {author} {\bibfnamefont {K.-F.}\ \bibnamefont {Chang}}, \bibinfo {author}
  {\bibfnamefont {H.-W.}\ \bibnamefont {Cho}}, \bibinfo {author} {\bibfnamefont
  {G.}~\bibnamefont {Juzeli{\=a}nas}}, \ and\ \bibinfo {author} {\bibfnamefont
  {A.~Y.}\ \bibnamefont {Ite}},\ }\href
  {https://www.nature.com/articles/ncomms6542} {\bibfield  {journal} {\bibinfo
  {journal} {Nature communications}\ }\textbf {\bibinfo {volume} {5}},\
  \bibinfo {pages} {5542} (\bibinfo {year} {2014})}\BibitemShut {NoStop}%
\bibitem [{\citenamefont {Wang}\ \emph
  {et~al.}(2017{\natexlab{a}})\citenamefont {Wang}, \citenamefont {Liu},
  \citenamefont {Xiao}, \citenamefont {Zhang}, \citenamefont {Alotaibi},
  \citenamefont {Sanders}, \citenamefont {Wang},\ and\ \citenamefont
  {Zhu}}]{wang2017strong}%
  \BibitemOpen
  \bibfield  {author} {\bibinfo {author} {\bibfnamefont {D.}~\bibnamefont
  {Wang}}, \bibinfo {author} {\bibfnamefont {C.}~\bibnamefont {Liu}}, \bibinfo
  {author} {\bibfnamefont {C.}~\bibnamefont {Xiao}}, \bibinfo {author}
  {\bibfnamefont {J.}~\bibnamefont {Zhang}}, \bibinfo {author} {\bibfnamefont
  {H.~M.~M.}\ \bibnamefont {Alotaibi}}, \bibinfo {author} {\bibfnamefont
  {B.~C.}\ \bibnamefont {Sanders}}, \bibinfo {author} {\bibfnamefont {L.-G.}\
  \bibnamefont {Wang}}, \ and\ \bibinfo {author} {\bibfnamefont
  {S.}~\bibnamefont {Zhu}},\ }\href
  {https://www.nature.com/articles/s41598-017-06266-0} {\bibfield  {journal}
  {\bibinfo  {journal} {Scientific reports}\ }\textbf {\bibinfo {volume} {7}},\
  \bibinfo {pages} {5796} (\bibinfo {year} {2017}{\natexlab{a}})}\BibitemShut
  {NoStop}%
\bibitem [{\citenamefont {Evers}\ and\ \citenamefont
  {Keitel}(2004)}]{evers2004double}%
  \BibitemOpen
  \bibfield  {author} {\bibinfo {author} {\bibfnamefont {J.}~\bibnamefont
  {Evers}}\ and\ \bibinfo {author} {\bibfnamefont {C.~H.}\ \bibnamefont
  {Keitel}},\ }\href
  {https://iopscience.iop.org/article/10.1209/epl/i2004-10207-5/fulltext/}
  {\bibfield  {journal} {\bibinfo  {journal} {EPL (Europhysics Letters)}\
  }\textbf {\bibinfo {volume} {68}},\ \bibinfo {pages} {370} (\bibinfo {year}
  {2004})}\BibitemShut {NoStop}%
\bibitem [{\citenamefont {Yi}\ \emph {et~al.}(2013)\citenamefont {Yi},
  \citenamefont {Gu},\ and\ \citenamefont {Li}}]{yi2013ground}%
  \BibitemOpen
  \bibfield  {author} {\bibinfo {author} {\bibfnamefont {Z.}~\bibnamefont
  {Yi}}, \bibinfo {author} {\bibfnamefont {W.-J.}\ \bibnamefont {Gu}}, \ and\
  \bibinfo {author} {\bibfnamefont {G.-X.}\ \bibnamefont {Li}},\ }\href
  {https://www.osapublishing.org/oe/abstract.cfm?uri=oe-21-3-3445} {\bibfield
  {journal} {\bibinfo  {journal} {Opt. Exp.}\ }\textbf {\bibinfo {volume}
  {21}},\ \bibinfo {pages} {3445} (\bibinfo {year} {2013})}\BibitemShut
  {NoStop}%
\bibitem [{\citenamefont {Huang}\ \emph {et~al.}(2016)\citenamefont {Huang},
  \citenamefont {Qi}, \citenamefont {Zhou}, \citenamefont {Niu},\ and\
  \citenamefont {Gong}}]{huang2016double}%
  \BibitemOpen
  \bibfield  {author} {\bibinfo {author} {\bibfnamefont {T.}~\bibnamefont
  {Huang}}, \bibinfo {author} {\bibfnamefont {Y.}~\bibnamefont {Qi}}, \bibinfo
  {author} {\bibfnamefont {F.}~\bibnamefont {Zhou}}, \bibinfo {author}
  {\bibfnamefont {Y.}~\bibnamefont {Niu}}, \ and\ \bibinfo {author}
  {\bibfnamefont {S.}~\bibnamefont {Gong}},\ }\href
  {https://www.sciencedirect.com/science/article/abs/pii/S0030402615019269}
  {\bibfield  {journal} {\bibinfo  {journal} {Optik-International Journal for
  Light and Electron Optics}\ }\textbf {\bibinfo {volume} {127}},\ \bibinfo
  {pages} {2978} (\bibinfo {year} {2016})}\BibitemShut {NoStop}%
\bibitem [{\citenamefont {Semerikov}\ \emph {et~al.}(2018)\citenamefont
  {Semerikov}, \citenamefont {Zalivako}, \citenamefont {Borisenko},
  \citenamefont {Khabarova},\ and\ \citenamefont
  {Kolachevsky}}]{semerikov2018eit}%
  \BibitemOpen
  \bibfield  {author} {\bibinfo {author} {\bibfnamefont {I.}~\bibnamefont
  {Semerikov}}, \bibinfo {author} {\bibfnamefont {I.}~\bibnamefont {Zalivako}},
  \bibinfo {author} {\bibfnamefont {A.}~\bibnamefont {Borisenko}}, \bibinfo
  {author} {\bibfnamefont {K.}~\bibnamefont {Khabarova}}, \ and\ \bibinfo
  {author} {\bibfnamefont {N.}~\bibnamefont {Kolachevsky}},\ }\href
  {https://link.springer.com/article/10.1007/s10946-018-9753-x} {\bibfield
  {journal} {\bibinfo  {journal} {Journal of Russian Laser Research}\ }\textbf
  {\bibinfo {volume} {39}},\ \bibinfo {pages} {568} (\bibinfo {year}
  {2018})}\BibitemShut {NoStop}%
\bibitem [{\citenamefont {Scharnhorst}\ \emph {et~al.}(2018)\citenamefont
  {Scharnhorst}, \citenamefont {Cerrillo}, \citenamefont {Kramer},
  \citenamefont {Leroux}, \citenamefont {W{\"u}bbena}, \citenamefont
  {Retzker},\ and\ \citenamefont {Schmidt}}]{scharnhorst2018experimental}%
  \BibitemOpen
  \bibfield  {author} {\bibinfo {author} {\bibfnamefont {N.}~\bibnamefont
  {Scharnhorst}}, \bibinfo {author} {\bibfnamefont {J.}~\bibnamefont
  {Cerrillo}}, \bibinfo {author} {\bibfnamefont {J.}~\bibnamefont {Kramer}},
  \bibinfo {author} {\bibfnamefont {I.~D.}\ \bibnamefont {Leroux}}, \bibinfo
  {author} {\bibfnamefont {J.~B.}\ \bibnamefont {W{\"u}bbena}}, \bibinfo
  {author} {\bibfnamefont {A.}~\bibnamefont {Retzker}}, \ and\ \bibinfo
  {author} {\bibfnamefont {P.~O.}\ \bibnamefont {Schmidt}},\ }\href@noop {}
  {\bibfield  {journal} {\bibinfo  {journal} {Physical Review A}\ }\textbf
  {\bibinfo {volume} {98}},\ \bibinfo {pages} {023424} (\bibinfo {year}
  {2018})}\BibitemShut {NoStop}%
\bibitem [{\citenamefont {Wang}\ \emph {et~al.}(2019)\citenamefont {Wang},
  \citenamefont {Qiao}, \citenamefont {Cai}, \citenamefont {Zhang},
  \citenamefont {Jin}, \citenamefont {Wang}, \citenamefont {Chen},
  \citenamefont {Luan}, \citenamefont {Wang}, \citenamefont {Song} \emph
  {et~al.}}]{wang2019realization}%
  \BibitemOpen
  \bibfield  {author} {\bibinfo {author} {\bibfnamefont {Y.}~\bibnamefont
  {Wang}}, \bibinfo {author} {\bibfnamefont {M.}~\bibnamefont {Qiao}}, \bibinfo
  {author} {\bibfnamefont {Z.}~\bibnamefont {Cai}}, \bibinfo {author}
  {\bibfnamefont {K.}~\bibnamefont {Zhang}}, \bibinfo {author} {\bibfnamefont
  {N.}~\bibnamefont {Jin}}, \bibinfo {author} {\bibfnamefont {P.}~\bibnamefont
  {Wang}}, \bibinfo {author} {\bibfnamefont {W.}~\bibnamefont {Chen}}, \bibinfo
  {author} {\bibfnamefont {C.}~\bibnamefont {Luan}}, \bibinfo {author}
  {\bibfnamefont {H.}~\bibnamefont {Wang}}, \bibinfo {author} {\bibfnamefont
  {Y.}~\bibnamefont {Song}},  \emph {et~al.},\ }\href
  {https://arxiv.org/abs/1912.04262} {\bibfield  {journal} {\bibinfo  {journal}
  {arXiv preprint arXiv:1912.04262}\ } (\bibinfo {year} {2019})}\BibitemShut
  {NoStop}%
\bibitem [{\citenamefont {Wang}\ \emph
  {et~al.}(2017{\natexlab{b}})\citenamefont {Wang}, \citenamefont {Um},
  \citenamefont {Zhang}, \citenamefont {An}, \citenamefont {Lyu}, \citenamefont
  {Zhang}, \citenamefont {Duan}, \citenamefont {Yum},\ and\ \citenamefont
  {Kim}}]{wang2017single}%
  \BibitemOpen
  \bibfield  {author} {\bibinfo {author} {\bibfnamefont {Y.}~\bibnamefont
  {Wang}}, \bibinfo {author} {\bibfnamefont {M.}~\bibnamefont {Um}}, \bibinfo
  {author} {\bibfnamefont {J.}~\bibnamefont {Zhang}}, \bibinfo {author}
  {\bibfnamefont {S.}~\bibnamefont {An}}, \bibinfo {author} {\bibfnamefont
  {M.}~\bibnamefont {Lyu}}, \bibinfo {author} {\bibfnamefont {J.-N.}\
  \bibnamefont {Zhang}}, \bibinfo {author} {\bibfnamefont {L.-M.}\ \bibnamefont
  {Duan}}, \bibinfo {author} {\bibfnamefont {D.}~\bibnamefont {Yum}}, \ and\
  \bibinfo {author} {\bibfnamefont {K.}~\bibnamefont {Kim}},\ }\href
  {https://www.nature.com/articles/s41566-017-0007-1} {\bibfield  {journal}
  {\bibinfo  {journal} {Nature Photonics}\ }\textbf {\bibinfo {volume} {11}},\
  \bibinfo {pages} {646} (\bibinfo {year} {2017}{\natexlab{b}})}\BibitemShut
  {NoStop}%
\bibitem [{\citenamefont {H{\"a}ffner}\ \emph {et~al.}(2003)\citenamefont
  {H{\"a}ffner}, \citenamefont {Gulde}, \citenamefont {Riebe}, \citenamefont
  {Lancaster}, \citenamefont {Becher}, \citenamefont {Eschner}, \citenamefont
  {Schmidt-Kaler},\ and\ \citenamefont {Blatt}}]{haffner2003precision}%
  \BibitemOpen
  \bibfield  {author} {\bibinfo {author} {\bibfnamefont {H.}~\bibnamefont
  {H{\"a}ffner}}, \bibinfo {author} {\bibfnamefont {S.}~\bibnamefont {Gulde}},
  \bibinfo {author} {\bibfnamefont {M.}~\bibnamefont {Riebe}}, \bibinfo
  {author} {\bibfnamefont {G.}~\bibnamefont {Lancaster}}, \bibinfo {author}
  {\bibfnamefont {C.}~\bibnamefont {Becher}}, \bibinfo {author} {\bibfnamefont
  {J.}~\bibnamefont {Eschner}}, \bibinfo {author} {\bibfnamefont
  {F.}~\bibnamefont {Schmidt-Kaler}}, \ and\ \bibinfo {author} {\bibfnamefont
  {R.}~\bibnamefont {Blatt}},\ }\href
  {https://journals.aps.org/prl/abstract/10.1103/PhysRevLett.90.143602}
  {\bibfield  {journal} {\bibinfo  {journal} {Phys. Rev. Lett.}\ }\textbf
  {\bibinfo {volume} {90}},\ \bibinfo {pages} {143602} (\bibinfo {year}
  {2003})}\BibitemShut {NoStop}%
\bibitem [{\citenamefont {Leibfried}\ \emph {et~al.}(2003)\citenamefont
  {Leibfried}, \citenamefont {Blatt}, \citenamefont {Monroe},\ and\
  \citenamefont {Wineland}}]{leibfried2003quantum}%
  \BibitemOpen
  \bibfield  {author} {\bibinfo {author} {\bibfnamefont {D.}~\bibnamefont
  {Leibfried}}, \bibinfo {author} {\bibfnamefont {R.}~\bibnamefont {Blatt}},
  \bibinfo {author} {\bibfnamefont {C.}~\bibnamefont {Monroe}}, \ and\ \bibinfo
  {author} {\bibfnamefont {D.}~\bibnamefont {Wineland}},\ }\href
  {https://journals.aps.org/rmp/abstract/10.1103/RevModPhys.75.281} {\bibfield
  {journal} {\bibinfo  {journal} {Reviews of Modern Physics}\ }\textbf
  {\bibinfo {volume} {75}},\ \bibinfo {pages} {281} (\bibinfo {year}
  {2003})}\BibitemShut {NoStop}%
\bibitem [{\citenamefont {Sawyer}\ \emph {et~al.}(2012)\citenamefont {Sawyer},
  \citenamefont {Britton}, \citenamefont {Keith}, \citenamefont {Wang},
  \citenamefont {Freericks}, \citenamefont {Uys}, \citenamefont {Biercuk},\
  and\ \citenamefont {Bollinger}}]{sawyer2012spectroscopy}%
  \BibitemOpen
  \bibfield  {author} {\bibinfo {author} {\bibfnamefont {B.~C.}\ \bibnamefont
  {Sawyer}}, \bibinfo {author} {\bibfnamefont {J.~W.}\ \bibnamefont {Britton}},
  \bibinfo {author} {\bibfnamefont {A.~C.}\ \bibnamefont {Keith}}, \bibinfo
  {author} {\bibfnamefont {C.~C.~Joseph}\ \bibnamefont {Wang}}, \bibinfo {author}
  {\bibfnamefont {J.~K.}\ \bibnamefont {Freericks}}, \bibinfo {author}
  {\bibfnamefont {H.}~\bibnamefont {Uys}}, \bibinfo {author} {\bibfnamefont
  {M.~J.}\ \bibnamefont {Biercuk}}, \ and\ \bibinfo {author} {\bibfnamefont
  {J.~J.}\ \bibnamefont {Bollinger}},\ }\href
  {https://journals.aps.org/prl/abstract/10.1103/PhysRevLett.108.213003}
  {\bibfield  {journal} {\bibinfo  {journal} {Phys. Rev. Lett.}\ }\textbf
  {\bibinfo {volume} {108}},\ \bibinfo {pages} {213003} (\bibinfo {year}
  {2012})}\BibitemShut {NoStop}%
\bibitem [{\citenamefont {Shankar}\ \emph {et~al.}(2019)\citenamefont
  {Shankar}, \citenamefont {Jordan}, \citenamefont {Gilmore}, \citenamefont
  {Safavi-Naini}, \citenamefont {Bollinger},\ and\ \citenamefont
  {Holland}}]{shankar2019modeling}%
  \BibitemOpen
  \bibfield  {author} {\bibinfo {author} {\bibfnamefont {A.}~\bibnamefont
  {Shankar}}, \bibinfo {author} {\bibfnamefont {E.}~\bibnamefont {Jordan}},
  \bibinfo {author} {\bibfnamefont {K.~A.}\ \bibnamefont {Gilmore}}, \bibinfo
  {author} {\bibfnamefont {A.}~\bibnamefont {Safavi-Naini}}, \bibinfo {author}
  {\bibfnamefont {J.~J.}\ \bibnamefont {Bollinger}}, \ and\ \bibinfo {author}
  {\bibfnamefont {M.~J.}\ \bibnamefont {Holland}},\ }\href
  {https://journals.aps.org/pra/abstract/10.1103/PhysRevA.99.023409} {\bibfield
   {journal} {\bibinfo  {journal} {Physical Review A}\ }\textbf {\bibinfo
  {volume} {99}},\ \bibinfo {pages} {023409} (\bibinfo {year}
  {2019})}\BibitemShut {NoStop}%
\bibitem [{\citenamefont {Johansson}\ \emph {et~al.}(2013)\citenamefont
  {Johansson}, \citenamefont {Nation},\ and\ \citenamefont
  {Nori}}]{johansson2013qutip}%
  \BibitemOpen
  \bibfield  {author} {\bibinfo {author} {\bibfnamefont {J.~R.}\ \bibnamefont
  {Johansson}}, \bibinfo {author} {\bibfnamefont {P.~D.}\ \bibnamefont
  {Nation}}, \ and\ \bibinfo {author} {\bibfnamefont {F.}~\bibnamefont
  {Nori}},\ }\href
  {https://www.sciencedirect.com/science/article/pii/S0010465512003955}
  {\bibfield  {journal} {\bibinfo  {journal} {Computer Physics Communications}\
  }\textbf {\bibinfo {volume} {184}},\ \bibinfo {pages} {1234} (\bibinfo {year}
  {2013})}\BibitemShut {NoStop}%
\bibitem [{\citenamefont {Lounis}\ and\ \citenamefont
  {Cohen-Tannoudji}(1992)}]{lounis1992coherent}%
  \BibitemOpen
  \bibfield  {author} {\bibinfo {author} {\bibfnamefont {B.}~\bibnamefont
  {Lounis}}\ and\ \bibinfo {author} {\bibfnamefont {C.}~\bibnamefont
  {Cohen-Tannoudji}},\ }\href
  {https://jp2.journaldephysique.org/articles/jp2/abs/1992/04/jp2v2p579/jp2v2p579.html}
  {\bibfield  {journal} {\bibinfo  {journal} {Journal de Physique II}\ }\textbf
  {\bibinfo {volume} {2}},\ \bibinfo {pages} {579} (\bibinfo {year}
  {1992})}\BibitemShut {NoStop}%
\bibitem [{\citenamefont {Hayes}\ \emph {et~al.}(2010)\citenamefont {Hayes},
  \citenamefont {Matsukevich}, \citenamefont {Maunz}, \citenamefont {Hucul},
  \citenamefont {Quraishi}, \citenamefont {Olmschenk}, \citenamefont
  {Campbell}, \citenamefont {Mizrahi}, \citenamefont {Senko},\ and\
  \citenamefont {Monroe}}]{hayes2010entanglement}%
  \BibitemOpen
  \bibfield  {author} {\bibinfo {author} {\bibfnamefont {D.}~\bibnamefont
  {Hayes}}, \bibinfo {author} {\bibfnamefont {D.~N.}\ \bibnamefont
  {Matsukevich}}, \bibinfo {author} {\bibfnamefont {P.}~\bibnamefont {Maunz}},
  \bibinfo {author} {\bibfnamefont {D.}~\bibnamefont {Hucul}}, \bibinfo
  {author} {\bibfnamefont {Q.}~\bibnamefont {Quraishi}}, \bibinfo {author}
  {\bibfnamefont {S.}~\bibnamefont {Olmschenk}}, \bibinfo {author}
  {\bibfnamefont {W.}~\bibnamefont {Campbell}}, \bibinfo {author}
  {\bibfnamefont {J.}~\bibnamefont {Mizrahi}}, \bibinfo {author} {\bibfnamefont
  {C.}~\bibnamefont {Senko}}, \ and\ \bibinfo {author} {\bibfnamefont
  {C.}~\bibnamefont {Monroe}},\ }\href
  {https://journals.aps.org/prl/abstract/10.1103/PhysRevLett.104.140501}
  {\bibfield  {journal} {\bibinfo  {journal} {Phys. Rev. Lett.}\ }\textbf
  {\bibinfo {volume} {104}},\ \bibinfo {pages} {140501} (\bibinfo {year}
  {2010})}\BibitemShut {NoStop}%
\bibitem [{\citenamefont {Turchette}\ \emph {et~al.}(2000)\citenamefont
  {Turchette}, \citenamefont {Kielpinski}, \citenamefont {King}, \citenamefont
  {Leibfried}, \citenamefont {Meekhof}, \citenamefont {Myatt}, \citenamefont
  {Rowe}, \citenamefont {Sackett}, \citenamefont {Wood}, \citenamefont {Itano},
  \citenamefont {Monroe},\ and\ \citenamefont
  {Wineland}}]{turchette2000heating}%
  \BibitemOpen
  \bibfield  {author} {\bibinfo {author} {\bibfnamefont {Q.~A.}\ \bibnamefont
  {Turchette}}, \bibinfo {author} {\bibfnamefont {D.}\ \bibnamefont {Kielpinski}}, \bibinfo
  {author} {\bibfnamefont {B.~E.}\ \bibnamefont {King}}, \bibinfo {author}
  {\bibfnamefont {D.}~\bibnamefont {Leibfried}}, \bibinfo {author}
  {\bibfnamefont {D.~M.}\ \bibnamefont {Meekhof}}, \bibinfo {author}
  {\bibfnamefont {C.~J.}\ \bibnamefont {Myatt}}, \bibinfo {author}
  {\bibfnamefont {M.~A.}\ \bibnamefont {Rowe}}, \bibinfo {author}
  {\bibfnamefont {C.~A.}\ \bibnamefont {Sackett}}, \bibinfo {author}
  {\bibfnamefont {C.~S.}\ \bibnamefont {Wood}}, \bibinfo {author}
  {\bibfnamefont {W.~M.}\ \bibnamefont {Itano}}, \bibinfo {author}
  {\bibfnamefont {C.}~\bibnamefont {Monroe}}, \ and\ \bibinfo {author}
  {\bibfnamefont {D.~J.}\ \bibnamefont {Wineland}},\ }\href {\doibase
  10.1103/PhysRevA.61.063418} {\bibfield  {journal} {\bibinfo  {journal} {Phys.
  Rev. A}\ }\textbf {\bibinfo {volume} {61}},\ \bibinfo {pages} {063418}
  (\bibinfo {year} {2000})}\BibitemShut {NoStop}%
\end{thebibliography}%


%

\onecolumngrid
\newpage
\section*{Appendix}

\section{The dark states}
Considering the semi-classical treatment for the system of Fig. 1(a) in the rotating frame, we have the Hamiltonian
\begin{equation}
\hat{H}=\begin{pmatrix}
0&\frac{\Omega_{\sigma-}}{2}&-\frac{\Omega_\pi}{2}&\frac{\Omega_{\sigma+}}{2}\\
\frac{\Omega_{\sigma-}}{2}&\Delta_{\rm d}+\delta_{\rm{B}}&0&0\\
-\frac{\Omega_\pi}{2}&0&\Delta_{\rm p}&0\\
\frac{\Omega_{\sigma+}}{2}&0&0&\Delta_{\rm d}-\delta_{\rm{B}}
\end{pmatrix}
\end{equation}
where the basis is $\{|e\rangle,\ket{+},|0\rangle,\ket{-}\}$, $\Delta_{\rm d}$ is the detuning between the driving laser and the $|0\rangle\leftrightarrow |e\rangle$ transition, $\Delta_{\rm p}$ is the detuning between the probe laser and the $|0\rangle\leftrightarrow|e\rangle$ transition, and $\delta_{\rm{B}}$ is the Zeeman splitting. Here we denote $\hbar=1$. 

Once the detuning of the probe beam matches to one of the Zeeman level $\Delta_{\rm p}=\Delta_{\rm d}+\delta_{\rm B}\equiv\Delta$, the Hamiltonian can be written as

\begin{equation}
\label{Hamiltonian}
\hat{H}=\begin{pmatrix}
0&\frac{\Omega_{\sigma-}}{2}&-\frac{\Omega_\pi}{2}&\frac{\Omega_{\sigma+}}{2}\\
\frac{\Omega_{\sigma-}}{2}&\Delta&0&0\\
-\frac{\Omega_\pi}{2}&0&\Delta&0\\
\frac{\Omega_{\sigma+}}{2}&0&0&\Delta-2\delta_{\rm{B}}
\end{pmatrix}
\end{equation}

This Hamiltonian gives us one dark state

\begin{equation}
    |D_1\rangle=\frac{1}{\sqrt{\Omega_\pi^2+\Omega_{\sigma-}^2}}(\Omega_\pi\ket{+}+\Omega_{\sigma-}|0\rangle)
\end{equation}

And the coincidence with the other Zeeman level gives us the second dark state

\begin{equation}
    |D_2\rangle=\frac{1}{\sqrt{\Omega_\pi^2+\Omega_{\sigma+}^2}}(\Omega_{\sigma+}|0\rangle+\Omega_\pi\ket{-})
\end{equation}

\newpage
\section{The scattering amplitude interpretation for the bright Resonance}
We can more precisely understand this four-level atomic system interacting with laser beams by quantizing laser field, which provides the Hamiltonian written as 

\begin{equation}
H=H_0+V_{\sigma+}+V_{\sigma-}+V_\pi+\sum_vV_v,
\end{equation}

where
\begin{equation}
\begin{aligned}
H_0&=  E_{\sigma+}\ket{-}\bra{-}+  E_{\pi}|0\rangle\langle0|+  E_{\sigma-}\ket{+}\bra{+} +   E_{e}|e\rangle\langle e|+ \omega_{\sigma-}a^\dagger_{\sigma-} a_{\sigma-}\\
&+ \omega_{\pi}a^\dagger_{\pi} a_{\pi}+ \omega_{\sigma+}a^\dagger_{\sigma+} a_{\sigma+}+\sum_v \omega_{v}a^\dagger_{v} a_{v}\\
V_{\sigma+}&=\frac{  d_{\sigma+}}{2}(a_{\sigma+}|e\rangle\bra{-}+a_{\sigma+}^\dagger\ket{-}\langle e|)\\
V_{\pi}&=\frac{  d_{\pi}}{2}(a_{\pi}|e\rangle\langle0|+a_{\pi}^\dagger|0\rangle\langle e|)\\
V_{\sigma-}&=\frac{  d_{\sigma-}}{2}(a_{\sigma-}|e\rangle\bra{+}+a_{\sigma-}^\dagger\ket{+}\langle e|)\\
V_{v}&=\frac{  d_{v}}{2}(a_{v}|e\rangle\bra{-}+a_{v}^\dagger\ket{-}\langle e|+a_{v}|e\rangle\langle0|+a_{v}^\dagger|0\rangle\langle e|+a_{v}|e\rangle\bra{+}+a_{v}^\dagger\ket{+}\langle e|).
\end{aligned}
\end{equation}
Here, the last term represent the interaction with the vacuum field. Basically, the absorption spectra is proportional to the squared scattering amplitude of the transition \begin{equation}
\ket{i}\equiv|0,1,N_1,N_2,0\rangle\to\ket{f}\equiv|0,0,N_1,N_2,1\rangle,
\end{equation}
where the first index represent the atom internal state, the second number represent the Fock state of the probe field, and the last three terms represent the Fock state of $\sigma_{-}$ field, $\sigma_{+}$ field and vacuum field. It is worthy to note that the absorption spectra means the atom absorb one photon and then emit it to vacuum, where part of it can be detected by PMT.

The scattering amplitude of such a process can be calculated by the $T$ matrix

\begin{equation}
T=\langle f|V| \mathfrak{i}\rangle+\lim _{\eta \rightarrow 0_{+}}\left\langle f\left|V \frac{1}{E_{\mathfrak{i}}-H+i \eta} V\right| i\right\rangle
\end{equation}

and due to $V=V_{\sigma+}+V_{\sigma-}+V_\pi+\sum_vV_v$, $V|i\rangle$ and $V|f\rangle$ can be written by

\begin{equation}
V|i\rangle= \frac{ \Omega_\pi}{2}|\varphi_e\rangle,~
V|f\rangle= \frac{ \Omega_v}{2}|\varphi_e\rangle.
\end{equation}
where we denote $|\varphi_e\rangle\equiv|e,0,N_1,N_2,0\rangle$. There are two states strongly coupled to $|\varphi_e\rangle$, which are 
\begin{equation}
|-,0,N_1+1,N_2,0\rangle,~ |+,0,N_1,N_2+1,0\rangle,  
\end{equation}
since 
\begin{equation}
V|+,0,N_1+1,N_2,0\rangle=\frac{ \Omega_{\sigma-}}{2}|\varphi_e\rangle,~ 
V|-,0,N_1,N_2+1,0\rangle=\frac{ \Omega_{\sigma+}}{2}|\varphi_e\rangle, 
\end{equation}
respectively. We note that the subspace is closed. We can calculate the $T$ matrix by projecting the Hamiltonian to the subspace spanned by $\{|+,0,N_1+1,N_2,0\rangle, |-,0,N_1,N_2+1,0\rangle,
|\varphi_e\rangle\}$. 

In the second order perturbation theory\cite{lounis1992coherent}, the effective Hamiltonian in this subspace can be calculated by 
\begin{equation}
P H_{\mathrm{eff}} P=P H_{0} P+P V P+P V Q \frac{1}{E_{0}-Q H_{0} Q} Q V P,
\end{equation}
where $P$ is the projection operator to the subspace and $Q=1-P$. After calculating all the terms, the effective Hamiltonian can be simplified by 
\begin{equation}
\hat{H}_{\rm{eff}}=\begin{pmatrix}
\Delta_\pi+i\frac{\Gamma}{2}&-\frac{\Omega_{\sigma-}}{2}&-\frac{\Omega_{\sigma+}}{2}\\
-\frac{\Omega_{\sigma-}}{2}&\Delta_\pi-\Delta_{\sigma-}&0\\
-\frac{\Omega_{\sigma+}}{2}&0&\Delta_\pi-\Delta_{\sigma+}. 
\end{pmatrix}
\end{equation}

Therefore the fluorescence, $W(\Delta)=|T(\Delta)|^2$, which is proportional to the square of the scattering amplitude is 

\begin{equation}
\label{eq:spectra}
W(\Delta_\pi)=\frac{16(\Delta_\pi-\Delta_{\sigma-})^2(\Delta_\pi-\Delta_{\sigma+})^2}{Z}
\end{equation}

where 
\begin{align*}
Z=4\Gamma^2(\Delta_\pi-\Delta_{\sigma-})^2(\Delta_\pi-\Delta_{\sigma+})^2+\left[4\Delta_\pi(\Delta_\pi-\Delta_{\sigma+})(\Delta_\pi-\Delta_{\sigma-})-(\Delta_\pi-\Delta_{\sigma+})\Omega_{\sigma-}^2-(\Delta_\pi-\Delta_{\sigma-})\Omega_{\sigma+}^2\right]^2
\end{align*}

And $\Gamma$ is the total decay rate of the excited state. The bright resonance will appear when

\begin{equation}
\label{eq:scattering_roots}
4\Delta_\pi(\Delta_\pi-\Delta_{\sigma-})(\Delta_\pi-\Delta_{\sigma+})-(\Delta_\pi-\Delta_{\sigma+})\Omega_{\sigma-}^2-(\Delta_\pi-\Delta_{\sigma-})\Omega_{\sigma+}^2=0
\end{equation}
In the experiment, we control $\Delta_\pi$ by changing the detuning of the probe beam that has only the $\pi$-polarization. 
The three roots independently correspond to the big Doppler peak and two narrow Fano peaks.

\begin{figure}[!htb]
\center{\includegraphics[width=0.6\textwidth]{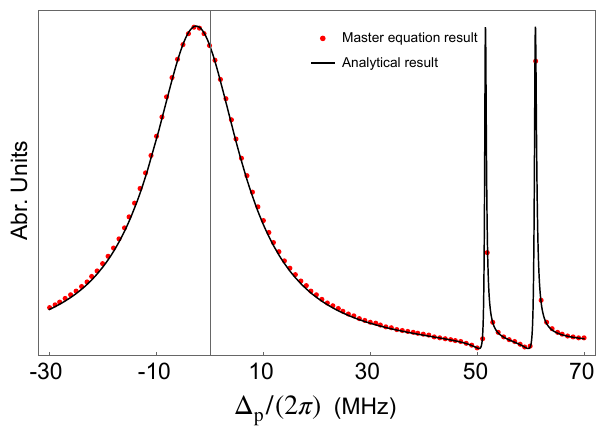}}
\caption{\label{fig:ODF_master}The spectrum calculated by the master equation and the analytical solution. Here we set $\Gamma/(2\pi)=21$ MHz, $\Delta_{\sigma+}/(2\pi)=50.4$ MHz, $\Delta{\sigma-}/(2\pi)=59.6$ MHz, $\Omega_{\sigma-}/(2\pi)=\Omega_{\sigma+}/(2\pi)=17$ MHz, and $\Omega_{\pi}/(2\pi)=0.5$ MHz. The dot represent the result calculated from the master equation and the curve is calculated by the Eq.(\ref{eq:spectra}).}
\end{figure}

\newpage
\section{The dressed-states interpretation for the bright resonance}
With only two driving beams and no probe beam, we have the Hamiltonian

\begin{equation}
\label{eq:dressed}
\hat{H}=\begin{pmatrix}
0&\frac{\Omega_{\sigma-}}{2}&0&\frac{\Omega_{\sigma+}}{2}\\
\frac{\Omega_{\sigma-}}{2}&\Delta_{\sigma-}&0&0\\
0&0&0&0\\
\frac{\Omega_{\sigma+}}{2}&0&0&\Delta_{\sigma+}
\end{pmatrix}
\end{equation}

The energy of the dressed states can be calculated by diagonalizing the above Hamiltonian, which results in solving the following equation

\begin{equation}
\frac{1}{4}\lambda\left[4\lambda(\lambda-\Delta_{\sigma+})(\lambda-\Delta_{\sigma-})-(\lambda-\Delta_{\sigma+})\Omega_{\sigma-}^2-(\lambda-\Delta_{\sigma-})\Omega_{\sigma+}^2\right]=0. 
\end{equation}



\begin{figure}[!htb]
\center{\includegraphics[width=0.7\textwidth]{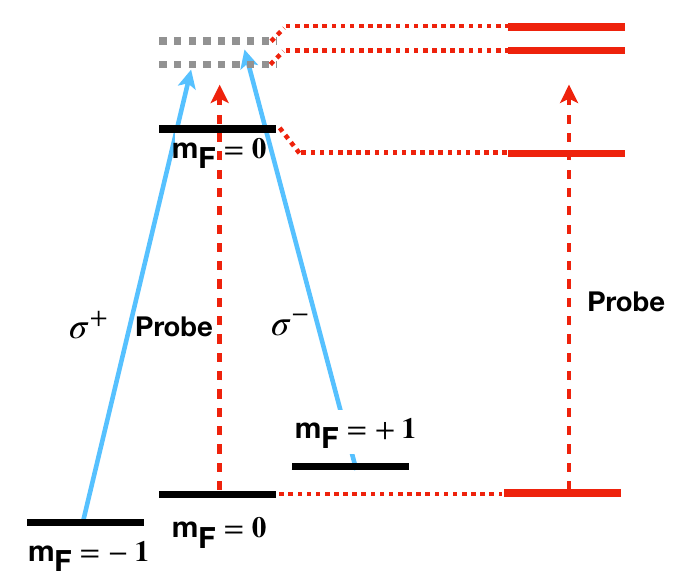}}
\caption{\label{fig:dressed}The dressed states. The gray dotted lines are the energy levels without the interaction with the $\Omega_{\sigma,\pm}$ fields. The gray solid lines are the dressed states formed by the laser field and the atomic levels. $|d_1,n\rangle$ and $|d_2,n\rangle$ represent the dressed states with phonon number $n$. $|g,n\rangle$ represent the $m_{\rm{F}}=0$ state with phonon number $n$. The dashed colored lines represent the virtual excited state of $|g\rangle$ with different phonon number, and blue- and red-sideband transitions, as shown in the spectrum. When the red dashed line matches $|d_2,n\rangle$, the transition $|g,n+1\rangle\leftrightarrow |d_2,n\rangle$ is driven, which is a red-sideband transition.}
\end{figure}

These energies of dressed states can be observed by applying a probe beam with $\pi$-polarization, as shown in Fig. \ref{fig:dressed}, which is consistent to the result of  Eq.(\ref{eq:scattering_roots}). 


\newpage
\section{The master equation treatment for the double-EIT cooling}
The Hamiltonian that describes the interaction between the four-level system and the laser fields shown in Fig. 1(a) can be written as

\begin{equation}
\hat{H}=\begin{pmatrix}
\omega_{\ket{e}}&\frac{\Omega_{\sigma-}}{2}e^{-i(\vec{k}_{\rm d}\cdot\vec{r}-\omega_{\rm d}t)}&-\frac{\Omega_\pi}{2}e^{-i(\vec{k}_{\rm p}\cdot\vec{r}-\omega_{\rm p}t)}&\frac{\Omega_{\sigma+}}{2}e^{-i(\vec{k}_{\rm d}\cdot\vec{r}-\omega_{\rm d}t)}\\
\frac{\Omega_{\sigma-}}{2}e^{i(\vec{k}_{\rm d}\cdot\vec{r}-\omega_{\rm d}t)}&\omega_{\ket{+}}&0&0\\
-\frac{\Omega_\pi}{2}e^{i(\vec{k}_{\rm p}\cdot\vec{r}-\omega_{\rm p}t)}&0&\omega_{\ket{0}}&0\\
\frac{\Omega_{\sigma+}}{2}e^{i(\vec{k}_{\rm d}\cdot\vec{r}-\omega_{\rm d}t)}&0&0&\omega_{\ket{-}},
\end{pmatrix}
\end{equation}
where $\vec{k}_{\rm p(d)}$ and $\omega_{\rm p(d)}$ are the k-vector and the frequency of the probe (driving) beam and $\omega_{\ket{e}}, (\omega_{\ket{-}}\omega_{\ket{0}},\omega_{\ket{+}})$ are the energies of the corresponding levels. For a rest ion, in the rotating frame the Hamiltonian can be simplified to 

\begin{equation}
\label{eq:sim}
\hat{H}_{\rm s}=\begin{pmatrix}
0&\frac{\Omega_{\sigma-}}{2}&-\frac{\Omega_\pi}{2}&\frac{\Omega_{\sigma+}}{2}\\
\frac{\Omega_{\sigma-}}{2}&\Delta_{\rm d}+\delta_{\rm{B}}&0&0\\
-\frac{\Omega_\pi}{2}&0&\Delta_{\rm p}&0\\
\frac{\Omega_{\sigma+}}{2}&0&0&\Delta_{\rm d}-\delta_{\rm{B}}
\end{pmatrix}
\end{equation}

The absorption spectrum can be obtained by numerical solving the steady state solution of the master equation corresponding to the Hamiltonian Eq.(\ref{eq:sim})

\begin{equation}
    \frac{d\hat\rho}{dt}=-i [\hat{H}_{\rm s},\hat \rho]+\mathcal{L}\rho,
\end{equation}
where $\mathcal{L}$ is the Lindblad operator corresponding to the three spontaneous decay channel  $\mathcal{L}\rho=\sum_{i=1}^3 c_i\rho c_i^\dagger-\frac{1}{2}\{c_i^\dagger c_i,\rho\}$ and $c_1=\sqrt{\Gamma/3}\ket{+}\bra{e}$, $c_2=\sqrt{\Gamma/3}\ket{0}\bra{e}$, $c_1=\sqrt{\Gamma/3}\ket{-}\bra{e}$. 

For a moving ion, the Hamiltonian in rotating frame can be written as

\begin{equation}
\label{eq:simplified}
\hat{H}_{\rm m}=\begin{pmatrix}
0&\frac{\Omega_{\sigma-}}{2}e^{-i\vec{k}_{\rm d}\cdot\vec{r}}&-\frac{\Omega_\pi}{2}e^{-i\vec{k}_{\rm p}\cdot\vec{r}}&\frac{\Omega_{\sigma+}}{2}e^{-i\vec{k}_{\rm d}\cdot\vec{r}}\\
\frac{\Omega_{\sigma-}}{2}e^{i\vec{k}_{\rm d}\cdot\vec{r}}&\Delta_{\rm d}+\delta_{\rm{B}}&0&0\\
-\frac{\Omega_\pi}{2}e^{i\vec{k}_{\rm p}\cdot\vec{r}}&0&\Delta_{\rm p}&0\\
\frac{\Omega_{\sigma+}}{2}e^{i\vec{k}_{\rm d}\cdot\vec{r}}&0&0&\Delta_{\rm d}-\delta_{\rm{B}}.
\end{pmatrix}
\end{equation}

In the simulation for the cooling effect, we set $\vec{k}_{\rm d}=-\vec{k}_{\rm p}=\hat{y}$ to deal only the relevant motional mode by the laser beams. For the quantized motion of ions trapped in the harmonic potential, the position operator can be decomposed to the creation and the annihilation operator of the phonon $\hat{y}=\sqrt{\frac{1}{2M\omega_{\rm COM}}}(\hat{a}+\hat{a}^\dagger)$, where $\omega_{\rm COM}$ is the frequency of the harmonic potential. The cooling speed can be calculated by solving the time evolution of the master equation corresponding to $\hat{H}_{\rm m}$ and the cooling limit can be calculated by the average phonon number of the steady state solution of the master equation.

The heating is represented by an additional Lindblad operator $c_{\rm h}=\alpha a^\dagger$ where $\alpha$ describes the strength of heating and $a^\dagger$ is the creation operator of the COM mode. For the numerical simulation in the main text, we use the parameter $\alpha=0.0259$, which is corresponding to a heating rate of $0.67$ quanta/ms, $\Delta_{\rm d}/(2\pi)=55.6$MHz, $\Delta_{\rm p}/(2\pi)=59.82$MHz, $\Omega_{\pi}/(2\pi)=6.67$MHz, $\Omega_{\sigma,+}/(2\pi)=19.91$MHz, $\Omega_{\sigma,-}/(2\pi)=19.52$MHz, $\delta_{\rm B}/(2\pi)=4.6$MHz and the trap frequency $\nu/(2\pi)=2.38$MHz. 
\newpage
\section{Measurements of Rabi Frequencies of the probe and driving beams}

\begin{figure}[!htb]
\center{\includegraphics[width=0.7\textwidth]{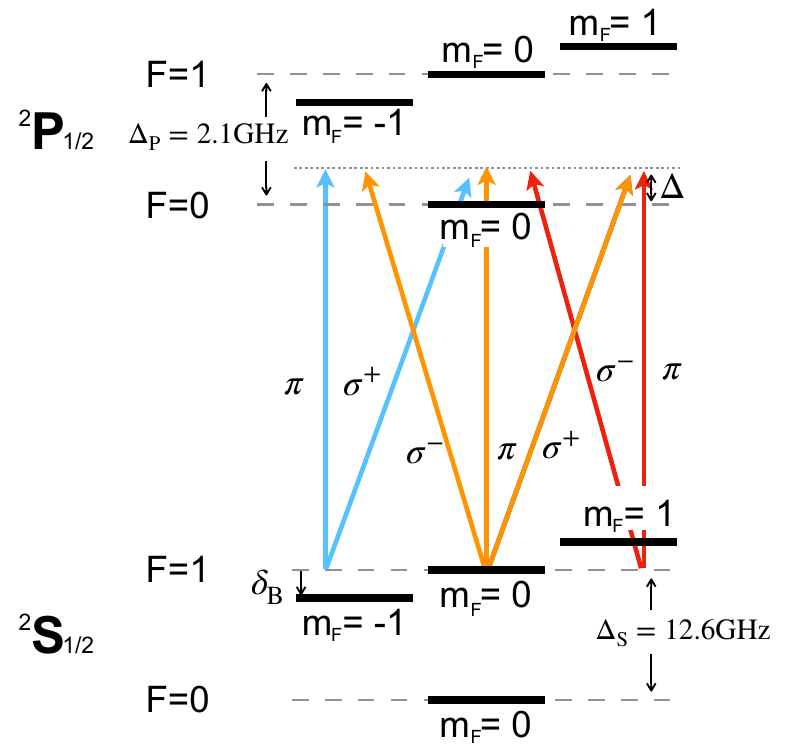}}
\caption{\label{fig:ac_stark}Energy levels of the \Yb. The AC stark shift originated from different transitions are labelled by different colors. The blue and red lines corresponding to the transitions contribute to the differential AC stark shift of the Zeeman qubits and the orange lines corresponding to the transitions for the clock qubit.}
\end{figure}

We measure the Rabi frequencies of the probe and driving laser beams by observing Ramsey oscillations from the differential AC-stark shift of the beams. We use the clock qubit and the Zeeman qubits to measure all the three components of polarization, as shown in the Fig.\ref{fig:ac_stark}. 

For the clock state qubit $^2S_{1/2}|F=0,m_F=0\rangle$ to $^2S_{1/2}|F=1,m_F=0\rangle$, the AC-stark is given by the following formula
\begin{equation}
\Delta^{\rm{clock}}_{\rm{AC}}(\Omega_+,\Omega_-,\Omega_\pi,\Delta)=\Omega_\pi^2\left(\frac{1}{\Delta}+\frac{1}{\Delta_{\rm P}+\Delta_{\rm S}-\Delta}\right)+(\Omega_-^2+\Omega_+^2)\left(\frac{1}{\Delta_{\rm P}+\Delta_{\rm S}-\Delta}-\frac{1}{\Delta_{\rm P}-\Delta}\right),
\end{equation}
where the first term comes from the transition $^2S_{1/2}|F=1,m_F=0\rangle$ to $^2P_{1/2}|F=0,m_F=0\rangle$ and $^2S_{1/2}|F=0,m_F=0\rangle$ to $^2P_{1/2}|F=1,m_F=0\rangle$ while the second term comes from the transition $^2S_{1/2}|F=1,m_F=0\rangle$ to $^2P_{1/2}|F=1,m_F=\pm1\rangle$ and $^2S_{1/2}|F=0,m_F=0\rangle$ to $^2P_{1/2}|F=1,m_F=\pm1\rangle$. Including the dephasing due to the spontaneous emission whose strength is proportional to the $1/\Delta^2$, the Ramsey oscillation can be described by 
\begin{equation}
\sin^2[\Delta^{\rm{clock}}_{\rm{AC}}(\Omega_+,\Omega_-,\Omega_\pi,\Delta)t]\times e^{-\Gamma*\Omega_\pi^2t/\Delta^2}e^{-\Gamma*(\Omega_-^2+\Omega_+^2)t/(\Delta_{\rm P}-\Delta)^2}.
\end{equation}
In a similar way, Ramsey oscillations of the two Zeeman qubits can be described by 
\begin{equation}
\sin^2[\Delta^{\pm1}_{\rm{AC}}(\Omega_+,\Omega_-,\Omega_\pi,\Delta)t]\times e^{-\gamma*\Omega_\mp^2t/(\Delta\pm\delta_{\rm B})^2}e^{-\gamma*\Omega_\pi^2t/(\Delta_{\rm P}-\Delta)^2},
\end{equation}
where $\Delta^{\pm1}_{\rm{AC}}$ are the differential AC-stark shifts of the Zeeman qubits which are given by

\begin{equation}
\begin{aligned}
\Delta^{\pm1}_{\rm{AC}}(\Omega_+,\Omega_-,\Omega_\pi,\Delta)&=\Omega_\mp^2\left(\frac{1}{\Delta\pm\delta_{\rm B}}-\frac{1}{\Delta_{\rm P}-\Delta\mp\delta_{\rm B}}+\frac{1}{\Delta_{\rm P}+\Delta_{\rm S}-\Delta}\right)+\\
&\Omega_\pi^2\left(-\frac{1}{\Delta_{\rm P}-\Delta}+\frac{1}{\Delta_{\rm P}+\Delta_{\rm S}-\Delta}\right)+\frac{\Omega_\pm^2}{\Delta_{\rm P}+\Delta_{\rm S}-\Delta}.
\end{aligned}
\end{equation}

\begin{figure}[!htb]
\center{\includegraphics[width=\textwidth]{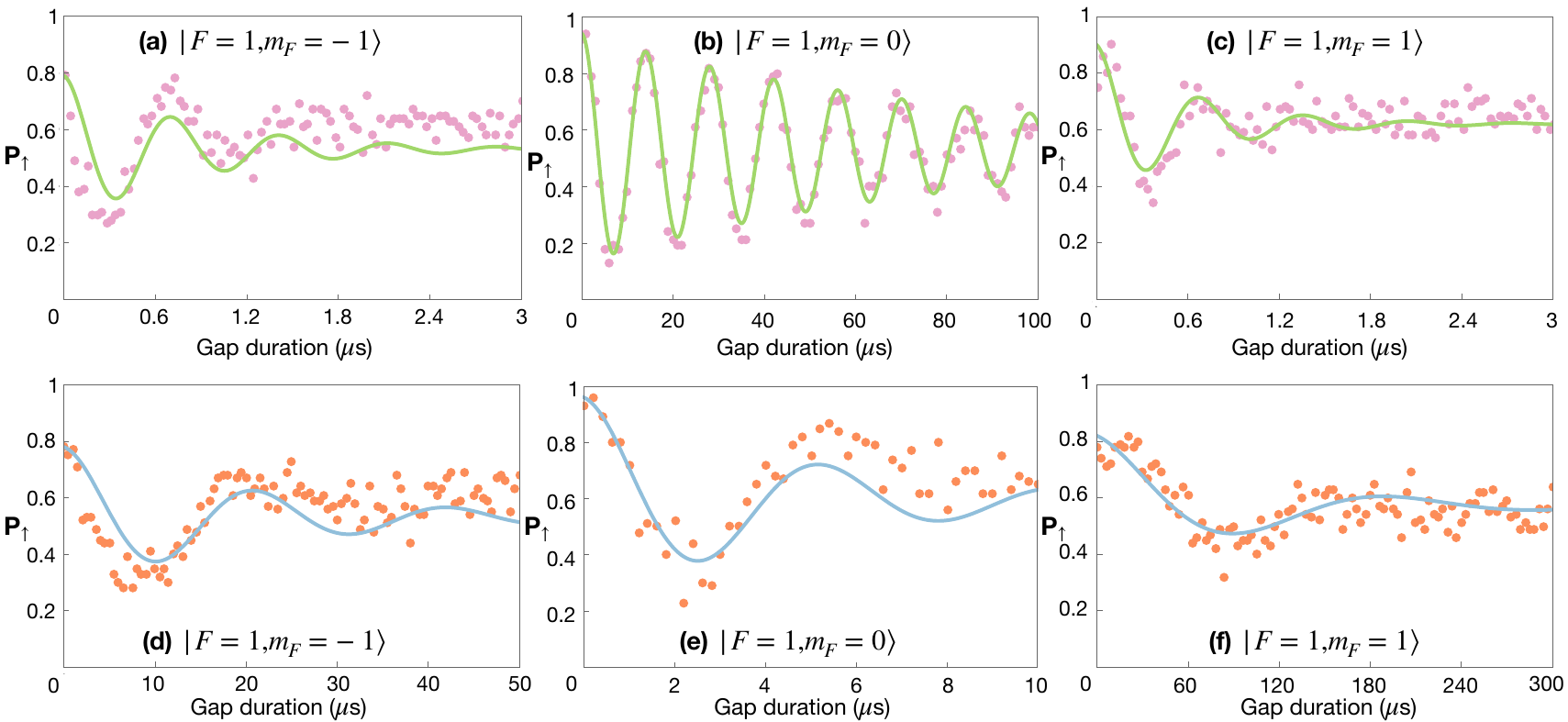}}
\caption{\label{fig:ramsey_result}Results of the Ramsey measurements (a-c) for the driving beam and (d-f) for the probe beam.}
\end{figure}

To measure the Ramsey oscillation, we first prepare the ion to its ground state $^2S_{1/2}|F=1,m_F=0\rangle$ by the optical pumping. Then a Ramsey sequence \cite{haffner2003precision} is applied. To measure the Rabi frequencies of the three components, we run the Ramsey sequence on all of three qubits, where measurement and fitting results are shown in Fig. \ref{fig:ramsey_result}. 

We note that the signal of Fig. \ref{fig:ramsey_result} (b) is used to align the direction of B-field. Once the B-field is parallel to the driving beam, the $\pi$ component of the driving beam will be eliminated, and the dephasing induced by the spontaneous emission will be reduced.

\newpage
\section{Extraction of the phonon number from the sideband transitions}
In our experiment the states $\ket{F=0,m_F=0}$ and $\ket{F=1,m_F=0}$ in the $^2S_{1/2}$ manifold, where the energy difference is $ \omega_{0}$, are defined as the $\ket{\downarrow}$ and $\ket{\uparrow}$ states of the qubit, respectively. A pair of 355 nm laser beams shown in Fig. 1(c) with a frequency difference $\omega_{\rm R}$ are used to drive the qubit through the Raman transition \cite{hayes2010entanglement}. As in the main text, we define $\mu_{\rm R}\equiv\omega_{\rm R}-\omega_{0}$ as the detuning of the Raman transition relative to the qubit transition.

When the detuning of the transition $\mu_{\rm R}$ matches the frequency of a motional mode, the Hamiltonians $\hat{H}_{\rm{r},m}$ and $\hat{H}_{\rm{b,m}}$, which represent the case of $\mu_{\rm{R}}=-\omega_m$ and $\mu_{\rm{R}}=\omega_m$, respectively,  can be written by

\begin{equation}
    \hat{H}_{{\rm r},m}=\hat{a}_m\sqrt{ \frac{1}{2M\omega_m}}\sum_{j}b_j^m\hat{\sigma}^{+}_j+\rm{h.c.}
\end{equation}
\begin{equation}
    \hat{H}_{{\rm b},m}=\hat{a}^\dagger_m\sqrt{ \frac{1}{2M\omega_m}}\sum_{j}b_j^m\hat{\sigma}^{+}_j+\rm{h.c.},
\end{equation}
where $M$ is the mass of single \Yb ion, $\hat{a}^\dagger_m$, $\hat{a}_m$ and $\omega_{m}$ are the creation, annihilation operator and angular frequency of the $m$-th motional mode, $b^m_j$ is the normal mode transformation matrix of the $j$-th ion with $m$-th mode. After the time evolution $\hat{U}_{{\rm r(b)},m}(t)$ of the Hamiltonian with an initial state $\ket{\downarrow\downarrow\downarrow\cdots\downarrow}\ket{n}_m$, where $\ket{n}_m$ is a Fock state of $m$-th mode, we can get the time dependence of the normalized average upstate population as 

\begin{equation}
    P_{\uparrow}^{{\rm r(b)},m}(t,n)=\operatorname{Tr}\left[\left(\sum_j \frac{\hat\sigma_{j}^{\rm z}+\hat{I}_{\rm s}}{2}\otimes\hat{I}_{m}\right)\rho_{{\rm r(b)},m}(t,n)\right]
\end{equation}
where $\hat{I}_{\rm s}$ and $\hat{I}_m$ are the identity operators of spins and the $m$-th mode and $\rho_{{\rm r(b)},m}(t,n)=\hat{U}(t)_{{\rm r(b)},m}\ket{\downarrow\downarrow\downarrow\cdots\downarrow}\bra{\downarrow\downarrow\downarrow\cdots\downarrow}\otimes\ket{n}_m\bra{n}_m\hat{U}(t)^{\dagger}_{{\rm r(b)},m}$ is the density matrix after the time evolution of duration $t$. If we start from a thermal state which is described by the density matrix
\begin{equation}
    \rho_{\rm {th},m}(\bar n) = \sum_i \frac{\bar n^i}{(\bar n+1)^{i+1}}\ket{i}_m\bra{i}_m.
\end{equation}

Instead of simulating the master equation with this density matrix as initial state, we numerically solve the time evolution from a Fock state with different phonon number and obtain the probabilities of a set $\{P_{\uparrow}^{{\rm r(b)},m}(t,0), P_{\uparrow}^{{\rm r(b)},m}(t,1), \cdots P_{\uparrow}^{{\rm r(b)},m}(t,n)\}$. Then the average upper-state probability after the time evolution from a thermal state can be calculated by the weighted superposition of each evolution of a Fock state, as shown below equation. 

\begin{equation}
    P_{\uparrow}^{{\rm r(b)},m}(t) =\operatorname{Tr}\left[\left(\sum_j \frac{\hat\sigma_{j}^{\rm z}+\hat{I}_{\rm s}}{2}\otimes\hat{I}_{m}\right)\rho_{\rm r(b), th,m}(t,\bar n)\right]=\sum_i \frac{\bar n^i}{(\bar n+1)^{i+1}}P_{\uparrow}^{{\rm r(b)},m}(t,i)
    \label{eq:averageup}
\end{equation}



At the $\pi$ duration of COM mode, we compare the ratio between blue-sideband and red-sideband transition for the $m$-th mode, $ P_{\uparrow}^{{\rm r},m}(t)/ P_{\uparrow}^{{\rm b},m}(t)$ to the Eq. (\ref{eq:averageup}). Then we deduce the $\bar{n}$ for the $m$-th mode. 

With this method we fit the peaks in Fig.4(b) and estimate the final temperature after the double-EIT cooling. The fitted results are \{0.101 (zig-zag mode), 0.0460, 0.0283, 0.0817, 0.0996, 0.0181, 0.0759, 0.0337, 0.0388, 0.0495, 0.0274, 1.04 (COM mode)\}.

It is interesting to discuss more about this method. For a single ion, we can extract the phonon number by calculating $\frac{P^{\rm r}/P^{\rm b}}{1-P^{\rm r}/P^{\rm b}}$ \cite{leibfried2003quantum}. However, this method is invalid if we globally drive the COM mode of the crystal. For instance, if we have a two ion crystal in a Fock state $|n=2\rangle$, the global red-sideband transition can only pump $50\%$ population of each ion to up state. Since the Hamiltonian of red-sideband transition is $\hat{H}=\eta\Omega_0\left[(\hat{\sigma}_{-}^{(1)}+\hat{\sigma}_{-}^{(2)})\hat{a}^\dagger+(\hat{\sigma}_{+}^{(1)}+\hat{\sigma}_{+}^{(2)})\hat{a}\right]$, we can reckon the red-sideband transition as an exchange between phonon and spin. Therefore, we can only excite one of spins by eliminating the phonon. The following figure Fig.\ref{fig:supply_global_red} gives an intuitive explanation

\begin{figure}[!htb]
\center{\includegraphics[width=0.8\textwidth]{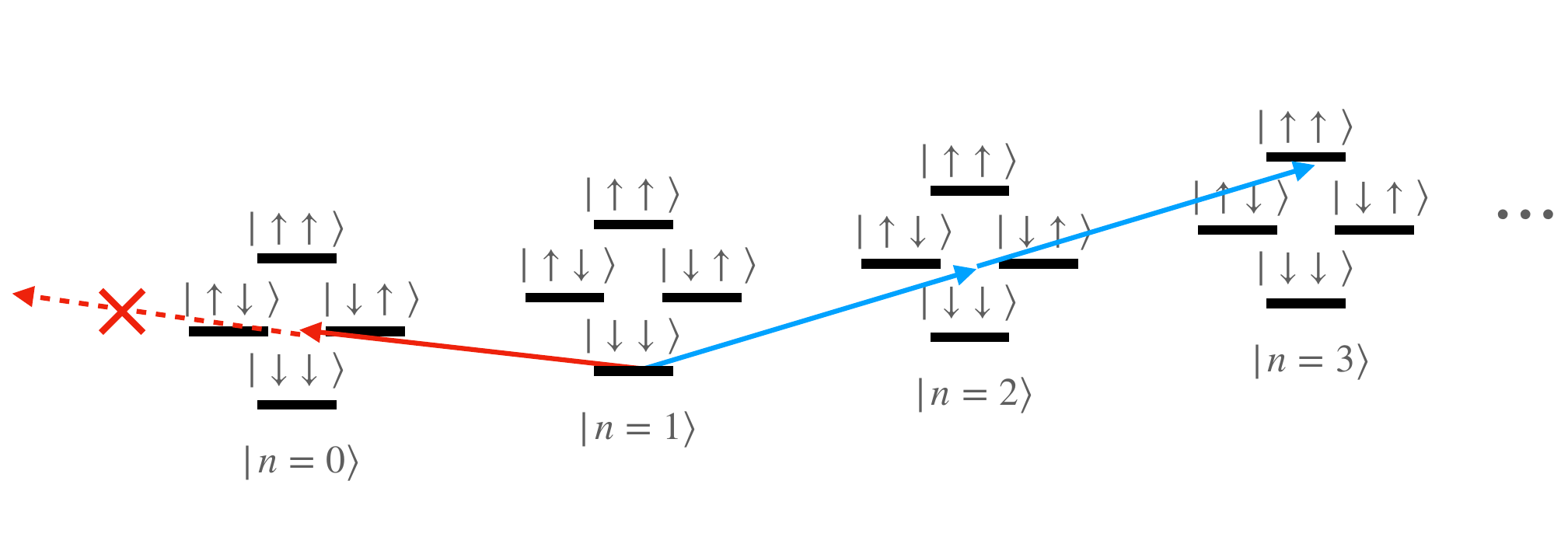}}
\caption{\label{fig:supply_global_red}Global red-sideband transition. Here, the red line represents the red-sideband transition, and the blue line represents the blue-sideband transition. Since the Fock state is bounded on $|n=0\rangle$, we can only reach $|\uparrow\downarrow\rangle$ or $|\downarrow\uparrow\rangle$ by globally driving the red-sideband transition of $|\downarrow\downarrow\rangle|n=1\rangle$. On the other hand, we can reach $|\uparrow\uparrow\rangle$ for the global blue-sideband transitions.}
\end{figure}

With this intuition, for 12 ions, we can also calculate the mean phonon number of COM mode by adding an additional parameter $\alpha$ to the formula $\bar{n} = \frac{\alpha P^{\rm r}/P^{\rm b}}{1-\alpha P^{\rm r}/P^{\rm b}}$. Fig. shows the relation between the ratio $P^{\rm r}/P^{\rm b}$ and $\bar{n}$, and indicates $\alpha\approx 2.7$ for $\bar n<2$. Here, we simulate the sideband transitions of COM mode with 12 ions with different initial phonon number up to $|n=10\rangle$ as discussed in Eq. (\ref{eq:averageup}). The Rabi frequency on the carrier transition is $\Omega/(2\pi)=12$kHz. The Lamb-Dicke parameter is $0.11/\sqrt{12}$. The transitions evolved for $206\mu $s to get the spectrum. We cut off the Fock space with phonon number larger than 22.

\begin{figure}[!htb]
\center{\includegraphics[width=0.6\textwidth]{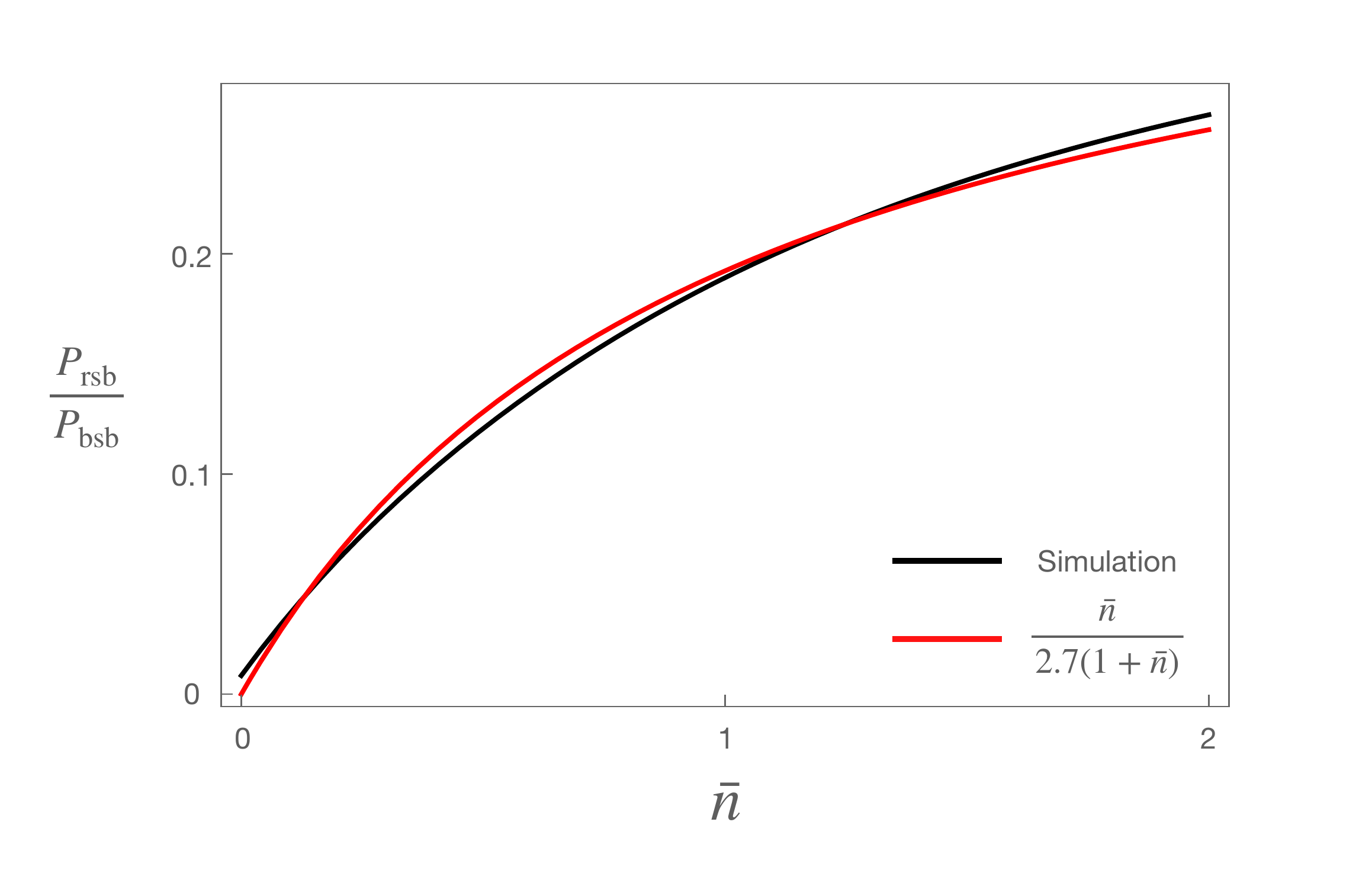}}
\caption{\label{fig:supply_PeakFit}Comparison between numerical simulation result and the formula $\frac{P^{\rm r}}{P^{\rm b}}=\frac{\bar n}{\alpha(1+\bar n)}$. Red curve represents the formula, and black curve represents the simulation result. Here $\alpha$ is $2.7$, and determined by fitting the simulation result. }
\end{figure}

\newpage
\section{Thermometry based on the optical-dipole-force}
\begin{figure}[!htb]
\center{\includegraphics[width=0.5\textwidth]{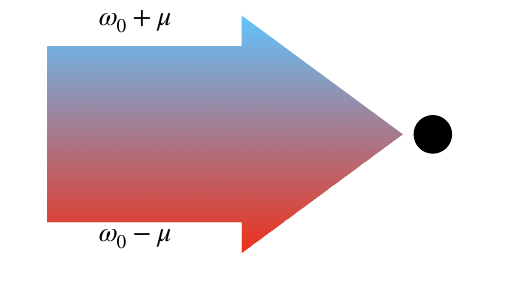}}
\caption{\label{fig:supply_ODF}Laser setting for the ODF measurement.}
\end{figure}

Fig. \ref{fig:supply_ODF} shows the laser setup for generating the optical-dipole-force(ODF). 

For the pair of Raman beam, the effective Hamiltonian can be written as

\begin{equation}
H_{I}^{(\mathrm{eff})}=\sum_j \frac{  \Omega_{j}}{2} e^{-i\left[\Delta k \cdot r_{j}(t)-\delta t-\Delta \varphi\right]}\hat\sigma_j^{+}+\mathrm{H.c.},
\end{equation}
where $\delta = \omega_{\rm R}-\omega_{0}$, $\Omega_{j}$ and $\Delta k$ are the Rabi frequency and net wave-vector of Raman laser beams, respectively. If we use two pair of Raman beam to generate two effective coupling simultaneously with opposite detuning $\mu_{\rm R}$ for the ODF, the whole Hamiltonian can be written as 
\begin{equation}
H_{I, \rm ODF}^{(\mathrm{eff})}=\sum_j\frac{  \Omega_{j}}{2} e^{-i\left[\Delta k_1 \cdot r_{j}(t)-\mu_{\rm R}t- \varphi_1\right]}\hat\sigma_j^{+}+\frac{  \Omega_{j}}{2} e^{-i\left[\Delta k_2 \cdot r_{j}(t)+\mu_{\rm R}t- \varphi_2\right]}\hat\sigma_j^{+}+\mathrm{H.c.}
\end{equation}

Though we have four transitions and two Raman transitions, we typically use only two laser beams. One has only one component, and the other has two frequency components. In this situation we have $\Delta k_1=\Delta k_2$. And we can rewrite the two phase terms by $\varphi_s=\frac{\varphi_1+\varphi_2}{2}$ and $\varphi_m=\frac{\varphi_1-\varphi_2}{2}$. Then the Hamiltonian becomes

\begin{equation}
H_{I, \rm ODF}^{(\mathrm{eff})}=\frac{  \Omega_{j}}{2}(\cos(\Delta k\cdot r_j)+i\sin(\Delta k\cdot r_j))\cos(\mu_{\rm R}t+\varphi_m)(\hat\sigma_x^{(j)}\cos\varphi_s-\hat\sigma_y^{(j)}\sin\varphi_s).
\end{equation}

\begin{figure}[!htb]
\center{\includegraphics[width=0.8\textwidth]{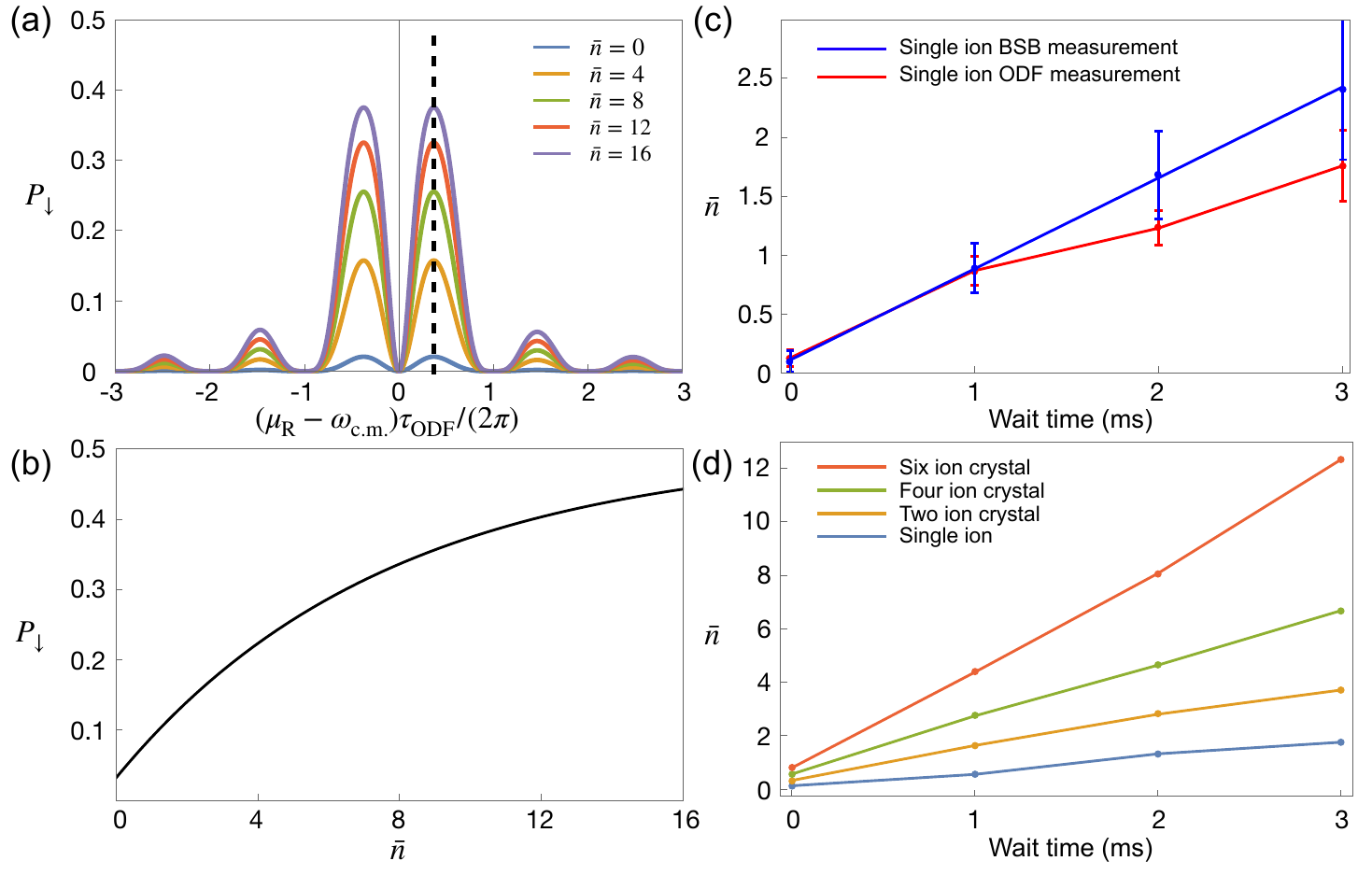}}
\caption{\label{fig:ODF_heating}ODF spectrum and the heating measurement. (a) The ODF spectrum of average up-state population with different phonon number for the COM mode. The dashed black line indicates the chosen detuning to measure the average phonon number. (b) The relation between the average phonon number and the average downstate population at the detuning of ODF pulse indicated in (a). Here the blue (red) area represents  error bars for the heating rate measured by sideband (ODF) method. (c) Comparison of the results of heating measurements between blue-sideband (blue line) and ODF height method (red line). (d) The heating measurement of multiple ions by the ODF height method. The measured heating rates are $0.77\pm0.19$ quanta/ms (sideband method) and $0.83\pm0.15$ quanta/ms (ODF method). (e) Linearly scaled heating rate. In this figure, the black line denotes $n\cdot \gamma_{\rm heating}$, where $n$ is the number of ions and $\gamma_{\rm heating}$ is the heating rate of a single ion.}
\end{figure}

In experiment, we calibrate the phase of two different frequency components to be the same, $\varphi_{s}=\varphi_{m}=0$, and the $\Delta k$ along the y direction. Then we have 

\begin{equation}
    H_{I}^{(\mathrm{eff})}=\frac{  \Omega_{j}}{2}\left[\cos(\Delta k\cdot y_j)\cos\mu_{\rm R}t+i\sin(\Delta k\cdot y_j)\cos\mu_{\rm R}t\right]\hat\sigma_x^{(j)},
\end{equation}
where the first term gives us the dephasing dependent on the motional state along the x-axis in the Bloch sphere \cite{sawyer2012spectroscopy}:

\begin{equation}
P_{\uparrow}^{j}=\frac{1}{2}\left[1-e^{-2 \Gamma_{\rm D} \tau} \exp \left(-2 \sum_{m}\left|\alpha_{j m}\right|^{2}\left(2 \bar{n}_{m}+1\right)\right)\right],
\end{equation}
where $\Gamma_{\rm D}$ describes the decoherence in the experiment and 

\begin{equation}
\label{eq:ODF_analytical}
\begin{aligned} \alpha_{j m}=& \eta_m{\Omega}_{j}\frac{ b_{j m}}{\left(\mu_{\rm R}^{2}-\omega_{m}^{2}\right)} \left(\omega_{m}(1-\cos \phi)+i \mu_{\rm R} \sin \phi\right.-e^{i \omega_{m} \tau}\left\{\omega_{m}\left[\cos \left(\mu_{\rm R} \tau\right)-\cos \left(\mu_{\rm R} \tau+\phi\right)\right]\right.\\ &\left.\left.-i \mu_{\rm R}\left[\sin \left(\mu_{\rm R} \tau\right)-\sin \left(\mu_{\rm R} \tau+\phi\right)\right]\right\}\right), \end{aligned}
\end{equation}
where $\phi=\left(\tau+\tau_{\pi}\right)\left(\mu_{\rm R}-\omega_{m}\right)$, $\eta_m = \Delta k \sqrt{\frac{1}{2 M \omega_{m}}}$ is the Lamb-Dicke parameter of m-th mode, and $\tau_\pi$ is the duration for the $\pi$-pulse during the ODF measurement. For the relative long duration of the ODF pulse we have $\phi\approx\tau\left(\mu_{\rm R}-\omega_{m}\right)$. When the detuning $\mu_{\rm R}$ is near the COM mode, we can approximate the Eq. (\ref{eq:ODF_analytical}) by

\begin{equation}
P_{\uparrow}^{j}=\frac{1}{2}\left[1-e^{-2 \Gamma_{\rm D} \tau} \exp \left(-2\left|\alpha_{j}\right|^{2}\left(2 \bar{n}+1\right)\right)\right],
\end{equation}

where 

\begin{equation}
\label{eq:ODF_analytical}
\begin{aligned} \alpha_{j}=& \frac{\eta{\Omega}_{j}}{\sqrt{N}}\frac{1}{\left(\mu_{\rm R}^{2}-\omega_{\rm{COM}}^{2}\right)} \left(\omega_{\rm{COM}}(1-\cos \phi)+i \mu_{\rm R} \sin \phi\right.-e^{i \omega_{\rm{COM}} \tau}\left\{\omega_{\rm{COM}}\left[\cos \left(\mu_{\rm R} \tau\right)-\cos \left(\mu_{\rm R} \tau+\phi\right)\right]\right.\\ &\left.\left.-i \mu_{\rm R}\left[\sin \left(\mu_{\rm R} \tau\right)-\sin \left(\mu_{\rm R} \tau+\phi\right)\right]\right\}\right), \end{aligned}
\end{equation}

where $\eta= \Delta k \sqrt{\frac{1}{2 M \omega_{\rm{COM}}}}$ is the Lamb-Dicke parameter for the COM mode, and $N$ is the number of ions. The spectrum resulting from the Eq.(\ref{eq:ODF_analytical}) with different phonon number are shown in \ref{fig:ODF_heating}(a). The null point corresponding to $\tau\left(\mu_{\rm R}-\omega_{m}\right)=2n\pi$. In the experiment we individually detect the fluorescence of ions in the crystal to get the downstate population and measure the strength of ODF by the Rabi oscillation of the carrier transition.

We develop a convenient way to study the cooling dynamics without obtaining the spectroscopy signal of Fig. \ref{fig:ODF_heating}(a) at each step of the cooling. We fix the detuning of the ODF beams at the highest peak of the spectrum shown in the dashed line of Fig. \ref{fig:ODF_heating}(a), which is $(\mu_{\rm R}-\omega_{\rm COM})\tau_{\rm ODF}/(2\pi)=0.37$ in our experiment, and record the up-state probability during the cooling. As shown in Fig.\ref{fig:ODF_heating}(a), with the same strength and duration of the ODF beams, the decrease of the temperature will lower the height of the spectrum. 

The relation between the up-state probability at the chosen detuning and the temperature is shown in the Fig.\ref{fig:ODF_heating}(b). By fitting the experimentally measured $P_{\uparrow}$ to the corresponding function we can quickly obtain the temperature of the mode, which we name as ODF height method. The reliability of this method is verified with the blue-sideband measurement for the heating of a single ion. We first measure the heating rate of a single ion by both blue-sideband method and the ODF height method. We compare their results, $0.83~ (\pm 0.19)~ \rm ms^{-1}$ for the blue-sideband method and $0.77~ (\pm 0.15)~ \rm ms^{-1}$ for the ODF method, which are consistent within error bars as shown in Fig.\ref{fig:ODF_heating}(c). Then we apply the ODF height method to measure the heating rate for COM mode of 2D crystals consisted of 1, 2, 4, and 6 ions. As shown in Fig.\ref{fig:ODF_heating}(d,e), the heating rate for the COM mode increases linearly with the number of ions as expected, which is $0.61~ (\pm 0.08)~ \rm ms^{-1}$ per ion. For an ion trapped by the electric fields, the heating rate can be calculated by \cite{turchette2000heating}

\begin{equation}
\dot{\bar{n}}=\frac{e^{2}}{4 m \hbar \omega}S_{E}\left(\omega\right)
\end{equation}

where $\dot{\bar{n}}$ is the heating rate, $\omega$ is the mode frequency, $\Omega_T$ is the frequency of the rf field, $m$ is the mass of the ion, $e$ is the elementary charge, and $S_E(\omega)\equiv 2\int_{-\infty}^\infty d\tau e^{i\omega\tau}\langle E(t)E(t+\tau)\rangle$ is the spectral density of electric-field fluctuations. The COM mode of a large crystal with $N$ ions can be considered as a single ion with mass $N m$ and charge $N e$. Then the coefficient $e^2/(4m\hbar\omega)$ will increase by a factor $N$, which indicates a linear dependence with the number of ions.

However, for the other modes of a large ion crystal, the above argument need some modifications. The Hamiltonian under a uniform electric noise is

$$
H(t)=\sum_i \hbar\omega_ia_i^\dagger a_i+eE(t)\sum_i x_i
$$

The ion's position operator can be decomposed into the normal coordinates $x_i=b_{ij}u_j$ and the normal coordinates $u_i=\sqrt{\frac{\hbar}{2m\omega_i}}(a_i+a_i^\dagger)$. Then the Hamiltonian can be simplified to

$$
H(t)=\sum_i \left[\hbar\omega_ia_i^\dagger a_i+u_ieE(t)\sum_jb_{ji} \right]
$$

\begin{figure}[!htb]
\center{\includegraphics[width=\textwidth]{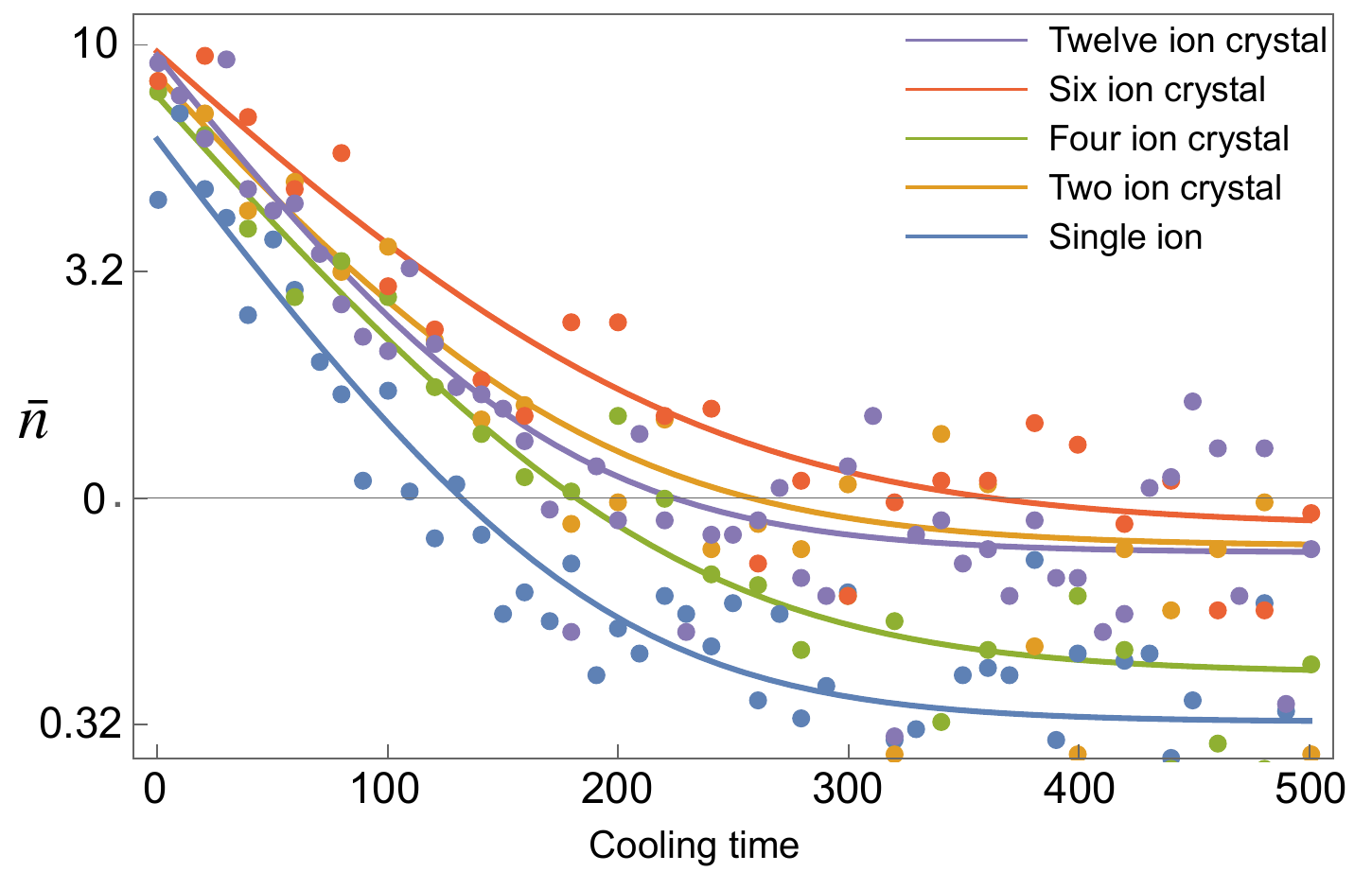}}
\caption{\label{fig:ODF_cooling}Multi-ion crystal cooling measurement by the ODF height method for 2, 4, 6 and 12 ions.}
\end{figure}

Based on the first-order perturbation theory \cite{turchette2000heating}, the heating rate on the i-th mode is

$$
\dot{\bar{n}}_i=\frac{e^2}{4m\hbar\omega_i}2\left(\sum_jb_{ji}\right)^2\int_{-\infty}^\infty d\tau e^{i\omega_i\tau}\langle E(t)E(t+\tau)\rangle=\left(\sum_jb_{ji}\right)^2\dot{\bar n}_{\rm single}
$$
\\
Again, for the COM mode, the heating rate of a large ion crystal will increase by a factor $(\sum_{i=1}^N\frac{1}{\sqrt{N}})^2=N$. The linear relation between the heating rate and the number of ions verifies the reliability of the ODF height measurement. We use it to probe the cooling speed on the crystals with a different number of ion. Figure \ref{fig:ODF_cooling} shows the cooling dynamics for the 2D crystal with different number of ions by using the ODF height method. Here, we note that the trap conditions are same to those with 12-ion crystal. The trap frequency of the transverse mode is almost two times lower than that for a single ion cooling, which results in higher cooling limit $\bar{n}=0.34\pm0.25$. For the twelve-ions, we averaged the experimental results over $390\mu$s to $500\mu$s, and got a mean phonon number of $1.04\pm 0.61$. The experimental data of single-, twelve-ion crystal and two-,four-,six-ion crystal are obtained at different experimental conditions, mostly the overall laser power of Doppler and EIT cooling beams, which causes inconsistency between them beyond the statistical errors. However, we can clearly conclude that there is no obvious speed-up of EIT cooling from many-body interaction. 
\newpage

\end{document}